\documentclass[sigconf,screen]{acmart}
\AtBeginDocument{%
  \providecommand\BibTeX{{%
    \normalfont B\kern-0.5em{\scshape i\kern-0.25em b}\kern-0.8em\TeX}}}

\copyrightyear{2024}
\acmYear{2024}
\setcopyright{rightsretained}
\acmConference[FAccT '24]{The 2024 ACM Conference on Fairness, Accountability, and Transparency}{June 3--6, 2024}{Rio de Janeiro, Brazil}
\acmBooktitle{The 2024 ACM Conference on Fairness, Accountability, and Transparency (FAccT '24), June 3--6, 2024, Rio de Janeiro, Brazil}\acmDOI{10.1145/3630106.3658941}
\acmISBN{979-8-4007-0450-5/24/06}

\acmSubmissionID{}

\usepackage{xcolor,colortbl}
\usepackage{tabularx}
\usepackage{multirow}
\usepackage{url}
\usepackage{subcaption}
\usepackage{color,soul}
\usepackage{enumitem}
\usepackage{hyperref}

\usepackage{arydshln}

\newcommand{\var}[1]{\texttt{#1}}
\newcommand{\cond}[1]{\textsc{#1}}

\usepackage[capitalize]{cleveref}
\crefname{section}{Section}{Sections}
\Crefname{section}{Section}{Sections}
\Crefname{table}{Table}{Tables}
\crefname{table}{Table}{Tables}
\crefname{figure}{Figure}{Figures}

\newcommand{\note}[1]{\textcolor{teal}{[#1]}}
\newcommand{\diff}[1]{\hl{#1}}

\begin{document}

\title[Examining the Impact of Large Language Models' Uncertainty Expression on User Reliance and Trust]{``I'm Not Sure, But...'': Examining the Impact of Large Language Models' Uncertainty Expression on User Reliance and Trust}

\author{Sunnie S. Y. Kim}
\authornote{Most work done during an internship at Microsoft.}
\email{sunniesuhyoung@princeton.edu}
\affiliation{%
  \institution{Princeton University}
  \country{USA}
}

\author{Q. Vera Liao}
\email{veraliao@microsoft.com}
\affiliation{%
  \institution{Microsoft}
  \country{Canada}
}

\author{Mihaela Vorvoreanu}
\email{mihaela.vorvoreanu@microsoft.com}
\affiliation{%
  \institution{Microsoft}
  \country{USA}
}

\author{Stephanie Ballard}
\email{stephballard@microsoft.com}
\affiliation{%
  \institution{Microsoft}
  \country{USA}
}

\author{Jennifer Wortman Vaughan}
\email{jenn@microsoft.com}
\affiliation{%
  \institution{Microsoft}
  \country{USA}
}

\makeatletter
\let\@authorsaddresses\@empty
\makeatother

\renewcommand{\shortauthors}{Kim, Liao, Vorvoreanu, Ballard, Vaughan}

\begin{abstract}
  Widely deployed large language models (LLMs) can produce convincing yet incorrect outputs, potentially misleading users who may rely on them as if they were correct. To reduce such overreliance, there have been calls for LLMs to communicate their uncertainty to end users. However, there has been little empirical work examining how users perceive and act upon LLMs' expressions of uncertainty. We explore this question through a large-scale, pre-registered, human-subject experiment (N=404) in which participants answer medical questions with or without access to responses from a fictional LLM-infused search engine. Using both behavioral and self-reported measures, we examine how different natural language expressions of uncertainty impact participants' reliance, trust, and overall task performance. We find that first-person expressions (e.g., ``\emph{I'm not sure, but...}'') decrease participants' confidence in the system and tendency to agree with the system's answers, while increasing participants' accuracy. An exploratory analysis suggests that this increase can be attributed to reduced (but not fully eliminated) overreliance on incorrect answers. While we observe similar effects for uncertainty expressed from a general perspective (e.g., ``\emph{It's not clear, but...}''), these effects are weaker and not statistically significant. Our findings suggest that using natural language expressions of uncertainty may be an effective approach for reducing overreliance on LLMs, but that the precise language used matters. This highlights the importance of user testing before deploying LLMs at scale.
\end{abstract}



\settopmatter{printacmref=false}
\maketitle

\noindent {\small \textbf{ACM Reference Format:}} \\
\noindent {\small Sunnie S. Y. Kim, Q. Vera Liao, Mihaela Vorvoreanu, Stephanie Ballard, and Jennifer Wortman Vaughan. 2024. ``I’m Not Sure, But...'': Examining the Impact of Large Language Models’ Uncertainty Expression on User Reliance and Trust. In \textit{The 2024 ACM Conference on Fairness, Accountability, and Transparency (FAccT ’24), June 3–6, 2024, Rio de Janeiro, Brazil.} ACM, New York, NY, USA. \url{https://doi.org/10.1145/3630106.3658941}}

\section{Introduction}
\label{sec:intro}

Large language models (LLMs) are transforming our daily lives. 
Today millions of people already incorporate LLMs into everyday tasks like searching for information~\cite{Mehdi_2023,liu2023evaluating}, writing~\cite{Gero2023CHI,Yuan2022IUI}, and programming~\cite{mozannar2023suggestion,al2022readable,pudari2023copilot}. 
However, the use of LLMs raises significant risks~\cite{bommasani2021opportunities,Bender2021StochasticParrots,Weidinger2022FAccT}.
Notably, like all models, LLMs are imperfect. They are widely recognized to produce outputs that are fluent and plausible, yet ultimately wrong~\cite{Ji2023Hallucination,truthfulqa,kabir2023answers}.
This can lead to disastrous outcomes through \emph{overreliance}~\cite{Bucinca2021CSCW,Vasconcelos2023CSCW,chen2023understanding,passi2022overreliance}, when people take actions based on incorrect outputs.
This concern garnered much public attention in 2023 when a lawyer included fake judicial opinions generated by ChatGPT in a legal brief presented in court~\cite{nytimes2023lawyer}. 
Such risks have been at the forefront of regulators' minds when drafting new frameworks for governing AI including the Draft AI Act in the European Union~\cite{EUAIAct} and the NIST AI Risk Management Framework in the United States~\cite{NISTAIRMF}.
In fact, Article 14 of the Draft EU AI Act explicitly requires developing and evaluating approaches to prevent overreliance on AI systems.
However, overreliance is notoriously difficult to mitigate, as many mitigations, such as explanations, are found to be ineffective or even can backfire to increase overreliance \cite{zhang2020effect,Bansal2021CHI,Poursabzi-Sangdeh-CHI2021,wang2021explanations,passi2022overreliance}.

As one approach to reduce overreliance, the research community has called for LLMs and LLM-infused applications to express the uncertainty of outputs to end users~\cite{vasconcelos2023generation,lin2022teaching,mielke-etal-2022-reducing,zhou2023navigating,baan2023uncertainty,liao2023ai}.
The idea of conveying AI uncertainty is not new; in AI-assisted decision-making settings, communicating (un)certainty has been shown to support trust calibration~\cite{zhang2020effect}, increase vigilance~\cite{Prabhudesai2023Uncertainty}, and improve task performance~\cite{Bansal2021CHI}.
But because of their open-ended outputs, wide-ranging use cases and user bases, and shifting public perception, LLMs raise new questions around how to both estimate and express uncertainty~\cite{vasconcelos2023generation,liao2023ai}.
For estimation, ``default'' approaches are often found to be overconfident~\cite{Chen2023Calibration,mielke-etal-2022-reducing,xiong2023llms}, and a new line of work has emerged on improving their calibration~\cite{mielke-etal-2022-reducing,Chen2023Calibration,kuhn2023semantic,zhou2023navigating,lin2022teaching,amayuelas2023knowledge,cole2023selectively,lin2023generating,tian2023just}.
For expression, LLMs open up a new design space; instead of presenting uncertainty numerically or visually, LLMs can present natural language expressions of uncertainty --- for instance, hedging phrases like ``\emph{I'm not sure, but...}'' --- embedded in their outputs.
Still, there is little understanding about how to effectively express uncertainty in natural language to end users. \looseness=-1

To deploy LLMs responsibly, it is necessary to understand how users react to uncertainty expression before implementing approaches at scale since it may have unintended negative consequences --- potentially even increasing overreliance if it causes the system to appear more trustworthy than it is.
Best practices for uncertainty expression will play a critical role in ensuring that requirements like those in the Draft EU AI Act serve their intended purpose.
To that end, we study how people perceive and act upon an LLM's expression of uncertainty when seeking medical information using a fictional LLM-infused search engine.
We choose to study this setting because search (unlike, for example, creative writing) is an application in which the correctness of responses is fundamental --- especially for potentially high-stakes medical queries --- making overreliance a serious concern.  
Additionally, LLM-infused search engines are already used by millions of people.\footnote{In March 2023, Microsoft reported Copilot in Bing served 45 million chats in the first month of its public preview~\cite{Mehdi_2023}. Perplexity AI reported its service had reached 2 million monthly active visitors in four months~\cite{Sullivan_2023,AI_2023}.} \looseness=-1

We choose to focus on natural language expressions for several reasons. 
First, LLM-infused search engines already include hedging language~\cite{mielke-etal-2022-reducing,liu2023evaluating,liao2023ai}.
Second, social science research shows that, in human communication, expressing (un)certainty through natural language is often preferred and perceived as more intuitive than numerical expressions~\cite{Dhami2022uncertainty,windschitl1996measuring,Liu2020VerbalNumerical,Wallsten1993PreferencesAR,Zimmer1983verbal}.
Third, this allows uncertainty to be expressed seamlessly within the natural language interactions of LLM-infused applications, rather than on the side or in onboarding materials~\cite{Cai2019HelloAI,ModelCards,openai2023gpt4} that users might overlook. \looseness=-1

Taking inspiration from the uncertainty communication literature --- both in the context of AI systems and person-to-person~\cite{oba2022hedges,Zhang2022CHI} --- we also explore the impact of the perspective used to express the uncertainty, comparing expressions in the first person (e.g., ``\emph{I'm not sure, but...}'') with expressions from a general perspective (e.g., ``\emph{It's not clear, but...}'').

Concretely, we conduct a large-scale, pre-registered, human-subject experiment (N=404) in which participants answer medical questions with or without access to responses from a fictional LLM-infused search engine, referred to as ``AI System A.''
We randomly vary whether participants have access to the system's responses as well as the presence (present/not present) and perspective (first-person/general) of uncertainty expressed in these responses.
We measure the impact of these experimental conditions on factors including participants' accuracy, the amount of time they take, their reliance on the system's responses versus other sources of information, and their self-reported trust in the system.

We find that participants who are shown first-person expressions of uncertainty are less confident in the system's answers, agree with the system's answers less often, and submit more correct answers compared with participants who see no expression of uncertainty.
An exploratory analysis suggests that the increased accuracy can be attributed to reduced (but not fully eliminated) overreliance on the system's incorrect answers.
While we observe similar effects for uncertainty expressed from a general perspective, these effects are weaker and not statistically significant. 
These results suggest that expressing uncertainty through natural language can be an effective way to reduce overreliance and (over)trust in LLM-infused search engines.
Still, we advocate for teams building and deploying LLMs to evaluate approaches to mitigate overreliance, including language choices, carefully with end users before release and for policymakers to embrace diverse and flexible approaches.

\begin{figure*}[t!]
\centering
\includegraphics[width=\textwidth]{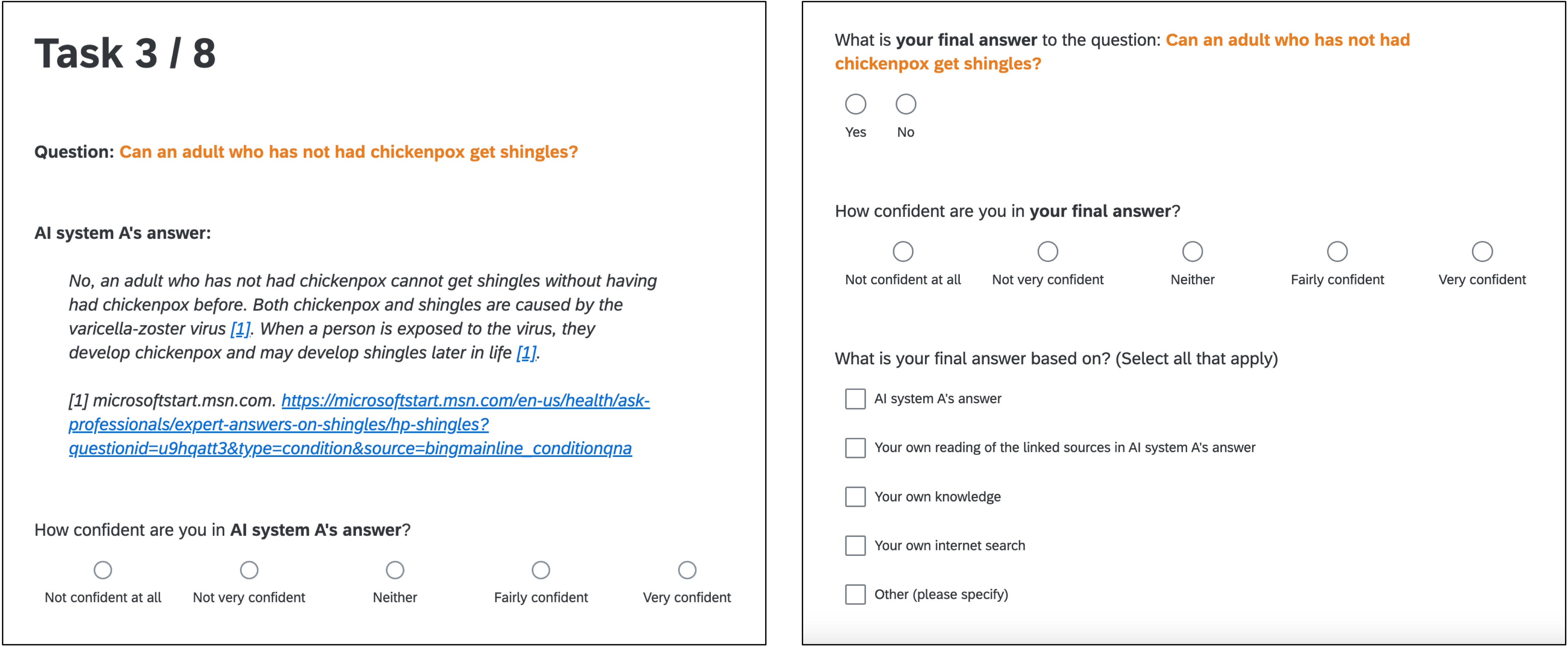}
\caption{\textbf{Screenshot of the information-seeking task} \textnormal{(shown in the \cond{Control}, \cond{Uncertain1st}, and \cond{UncertainGeneral} conditions). Additional screenshots, including the variant shown in the \cond{No-AI} condition, can be found in Appendix D.}}
\Description{The screenshot displays a task question, the AI system's answer, and a set of questions to be answered by participants. See Appendix D for the full wording and Appendix E for the task questions and AI responses used in the experiment.}
\label{fig:introduction}
\end{figure*}

\section{Related Work}
\label{sec:relatedwork}

\subsection{Uncertainty Expression}

Uncertainty expression has been studied extensively both in the context of AI and in the context of human communication.
Estimates of uncertainty can be expressed in different ways, including numerically (e.g., ``\emph{a probability of 0.2}''), visually (e.g., displaying error bars), and through natural language (e.g., ``\emph{with high uncertainty...}'').
While numerical expressions and visualizations allow for high precision, they are notoriously difficult for people to understand and are often misinterpreted, even by experts~\cite{Kahneman2013thinking,sunstein2002probability,hofman2020visualizing,belia2005researchers}.
In contrast, while less precise, natural language expressions of uncertainty are often perceived to be more intuitive and favored by people~\cite{Dhami2022uncertainty,windschitl1996measuring,Liu2020VerbalNumerical,Wallsten1993PreferencesAR,Zimmer1983verbal}.
Because of this, and since LLMs already produce natural language outputs, we focus on natural language expressions in our study.

Different forms of natural language uncertainty expression have been studied by researchers in disciplines ranging from psychology~\cite{Clark1990Verbal,vogel2022interpretation,Wesson2009Review} and human-computer interaction~\cite{Følstad2020ChatbotRepair,Ashktorab2019ChatbotRepair,Torrey2013HRI} to communication~\cite{Lorson2021Hedging} and marketing~\cite{oba2022hedges,Durik2008hedges}. Our study design builds on this literature.
Most notably, in the context of marketing, Oba and Berger~\cite{oba2022hedges} found that different types of hedges --- a form of uncertainty expression --- have different levels of persuasion, with the most persuasive being those that suggest a high likelihood of occurrence (e.g., ``\emph{probably}'' as opposed to ``\emph{possibly}'') and those that take a personal, first-person perspective (e.g., ``\emph{I feel like...}'') as opposed to a general perspective (e.g., ``\emph{It feels like...}'').
Indeed, taking a first-person perspective is often found to increase the engagement in the persuasion literature~\cite{chen2017persuasive}.
This motivated our exploration of the effect of AI uncertainty expressions taking a first-person or general perspective, though we explore the \textit{negative} persuasive effect (i.e., deterring overreliance) of expressing uncertainty.

Many methods have been proposed for estimating and communicating uncertainty in AI systems~\cite{Bhatt2021Uncertainty}.
Most relevant, several studies have shown that communicating uncertainty can reduce overreliance.
In the context of house valuation, \citet{Prabhudesai2023Uncertainty} found that visualizing a system's uncertainty using quantile dot plots forced participants to slow down and think analytically.
In the context of sentiment classification and question answering, \citet{Bansal2021CHI} found that displaying a system's numerical confidence improved participants' task performance. \looseness=-1

A few studies have looked specifically at natural language expressions of uncertainty in AI systems. They have been studied in the chatbot literature as a design strategy to prevent user frustration from conversational breakdowns~\cite{Ashktorab2019ChatbotRepair,Følstad2020ChatbotRepair}. 
\citet{Radensky2023FAccT} studied communicating the confidence of a conversational music recommendation system through language and color coding.
Experimenting with different levels of calibration, they found that accurate confidence signals led to the greatest increase in trust without encouraging overreliance, though underreliance may still occur. 
\citet{Zhang2022CHI} explored different ways of having an AI system (named ``ShapeBot'') express confidence in the reasoning behind its recommendation, examining the effect of point of view (first-person ``\emph{I think...}'' vs. third-person ``\emph{ShapeBot thinks...}'') and strength of the belief expressed (``\emph{ShapeBot thinks...}'' vs. ``\emph{ShapeBot knows...}'').
They found that both factors affected user reliance, highlighting the importance of carefully considering the language used to express (un)certainty. \looseness=-1

\subsection{Uncertainty in LLMs}

Obtaining accurate numerical estimates of uncertainty for LLMs is an active line of research.
One way to estimate an LLM's uncertainty is by the likelihood of generating a specific output given the context.
However, this ``generation probability'' may not reflect what end users expect or want when they think of uncertainty~\cite{vasconcelos2023generation}.
A more useful notion of uncertainty might be one that captures how likely it is that the LLM's output is factually correct or correctly meets the user's needs.
This notion of uncertainty may apply to a full output or to sentences, phrases, or words within the output.
Many researchers are working on evaluating how calibrated existing uncertainty estimates are and proposing new techniques to improve calibration~\cite{mielke-etal-2022-reducing,Chen2023Calibration,kuhn2023semantic,zhou2023navigating,lin2022teaching,amayuelas2023knowledge,cole2023selectively,lin2023generating,tian2023just}.
Current findings suggest that LLMs are often overconfident~\cite{Chen2023Calibration,mielke-etal-2022-reducing,xiong2023llms}, which may give a false impression of their capabilities and exacerbate overreliance~\cite{dinan2022safetykit}. \looseness=-1

There is also a growing interest in LLMs' ability to directly generate natural language expressions of (un)certainty~\cite{xiong2023llms,lin2022teaching,mielke-etal-2022-reducing,zhou2023navigating}.
Notably, \citet{mielke-etal-2022-reducing} observed that LLMs regularly express confidence (e.g., ``\emph{Obviously...}'') and doubt (e.g., ``\emph{I'm not sure, but...}'') through the language used in their outputs, but these expressions are poorly calibrated.
\citet{zhou2023navigating} ``taught'' OpenAI's GPT-3~\cite{GPT3NeurIPS2020} model to express (un)certainty through prompt engineering, but also found that the generated expressions were not well calibrated, especially those suggesting high certainty.

Despite this active research, there has been little empirical work examining the impact of uncertainty expression on users of LLM-infused systems. 
Notable exceptions are the works of \citet{vasconcelos2023generation} and \citet{spatharioti2023comparing}, who explored the effect of highlighting uncertain parts of LLM outputs in the context of code completion and search, respectively, and the concurrent work of \citet{zhou2024relying}, who explored the effect of LLMs' natural language expressions of (un)certainty in the context of trivia question answering. 
These studies' results support uncertainty expression as a promising technique to encourage appropriate reliance, particularly when uncertainty estimates are well calibrated.
Our work adds empirical knowledge on this topic through a large-scale, pre-registered experiment studying natural language uncertainty expressions in the context of LLM-infused search.

To avoid making assumptions of calibration or tying our experiment to a particular uncertainty estimation approach, we design our study to include both instances in which the system expresses uncertainty when it is incorrect and instances in which it expresses uncertainty when it is correct. 
By randomly varying whether or not uncertainty is expressed on any particular response, we are able to directly compare participants' behavior when uncertainty is and is not expressed. \looseness=-1

\subsection{Measuring Reliance and Trust}
\label{sec:measuring}

We hypothesize that whether or not an AI system expresses uncertainty --- and if it does, the perspective in which the uncertainty is expressed --- impacts user reliance and trust.
We note that there are many definitions, measures, and factors of reliance and trust~\cite{Ueno2022CHIEA,Vereschak2021CSCW,Kim2023FAccT,Zhang2023CHI,Cheng2021Trust}.
We use a combination of dependent variables and a mix of behavioral and self-reported measures to capture aspects most relevant to our research setting of LLM-infused search.

In the AI-assisted decision-making literature, many experiments follow the judge-advisor paradigm~\cite{bonaccio2006advice}, presenting a participant with an AI-generated answer to a question and then asking the participant to provide their own answer, a set-up we adopt in our work. 
In this set-up, \emph{agreement} between a participant's answer and that of the AI system is a commonly studied behavioral measure of reliance and trust~\cite{yin2019understanding,zhang2020effect,Lai2019FAccT,Bucinca2021CSCW,Cao2022CSCW,Liu2021CSCW,Lu2021CHI,Mohseni2020}.
While this does not capture reliance or trust directly --- the participant may have come up with the same answer on their own even without the AI system --- comparing how often participants agree with the AI system's answers across experimental conditions gives a way of measuring whether they rely on the system differently across conditions.
We use this measure in our work.
We note that in some prior work, the participant is asked to provide an initial answer first before seeing that of the AI system~\cite{Poursabzi-Sangdeh-CHI2021,logg17theory,logg19people,Schemmer2023IUI}.
In these cases, other metrics like weight of advice can be used to more directly capture reliance.
We do not adopt this set-up because we use questions we do not expect participants to be able to answer on their own. \looseness=-1

To complement \emph{agreement}, we examine participants' \emph{confidence}, \emph{source usage}, \emph{trust intentions}, and \emph{trust beliefs}.
First, as in prior work~\cite{Poursabzi-Sangdeh-CHI2021,Cao2022CSCW,Kim2022HIVE,CHONG2022,Lu2021CHI}, we ask participants to report their \emph{confidence} both in the answer output by the AI system and in their own answer. 
Second, as an indirect measure of reliance and trust, we capture participants' \emph{source usage} by both tracking whether or not participants click on the linked sources in the system's responses and asking them to self-report the resources that they based their final answer on.
Finally, using responses collected in an exit questionnaire, we measure participants' \emph{trust intentions} and \emph{trust beliefs} using the scales developed by McKnight et al.~\cite{McKnight2002}.
Trust intentions refer to a participant's desire to use the system, while trust beliefs refer to their perceptions about the system's trustworthiness such as the system's perceived ability, benevolence, and integrity~\cite{Mayer1995trust}. 
In general, the two are positively related~\cite{McKnight2002}, but \citet{Radensky2023FAccT} found that they can be differently impacted by a system's expressed confidence.
We measure both in our experiment to better understand the impact of the system's uncertainty expression. \looseness=-1

We also measure two system facets that are known to impact trust: \emph{perceived anthropomorphism} and \emph{perceived transparency}. 
Participants may view the expression of uncertainty (especially first-person) as an inherently human behavior, leading to increased anthropomorphism.
Recent work has expressed concern around anthropomorphism leading to over-trust~\cite{abercrombie2023mirages,shelby2022sociotechnical}, a potential path for uncertainty expression to backfire.
Uncertainty expression can also increase the system's perceived transparency~\cite{Bhatt2021Uncertainty}, which is generally shown to enhance trust, whether or not appropriate~\cite{liao2023ai}. \looseness=-1

In addition to reliance and trust, we also consider task performance as a dependent variable, measured as \emph{correctness} of participants’ answers and \emph{time on task}.
Both of these have been studied in prior work on AI-assisted decision making~\cite{lai2021towards}, as well as specifically for studying the effect of uncertainty expression~\cite{Vasconcelos2023CSCW}.

We formally define the dependent variables in Section~\ref{sec:dvs} and articulate hypotheses in Section~\ref{sec:analysis}.

\section{Methods}
\label{sec:methods}

As described in Section~\ref{sec:measuring}, our experiment is designed to measure the impact of natural language expressions of an LLM's uncertainty on user reliance and trust.
We do this in the context of information seeking in the medical domain.
We pre-registered our experimental design, hypotheses, analysis plan, and data collection procedures before collecting data.\footnote{Our pre-registration is viewable at \url{https://osf.io/mnrp9}.}
To complement our pre-registered analyses, we include exploratory analyses and a qualitative analysis of participants' free-form responses.
The study was approved by our internal Institutional Review Board (IRB). \looseness=-1

\subsection{Procedure and Experimental Conditions}

We designed a between-subjects experiment with some within-subjects comparisons, which we conducted on Qualtrics.
Participants complete a set of information-seeking tasks. 
Each task involves determining the correct yes-or-no answer to a challenging, factual question in the medical domain with or without access to responses from a fictional LLM-infused search engine, ``AI system A.''
The presence and form of system responses provided to participants depend on their experimental condition.
Specifically, participants are randomly placed into one of four experimental conditions: 
\begin{itemize}[topsep=0pt]
    \item \cond{Control}: Participants see AI responses without any expression of uncertainty. 
    \item \cond{Uncertain1st}: Participants see AI responses and half of the time these responses include uncertainty expressed in the first person, with personal pronouns (e.g., ``\emph{I’m not sure, but it seems...}''). 
    \item \cond{UncertainGeneral}: Participants see AI responses and half of the time these responses include uncertainty expressed in a general perspective, without personal pronouns (e.g., ``\emph{There is uncertainty, but it seems...}''). 
    \item \cond{No-AI}: Participants are not told about the AI system and do not see AI responses.
\end{itemize}
\cond{Control} is a baseline to which we compare the conditions \cond{Uncertain1st} and \cond{UncertainGeneral} to understand the impact of uncertainty expressions.
\cond{No-AI} is a second baseline to understand the impact of access to the AI system. \looseness=-1

The experiment is divided into three components.
In the first, participants are introduced to the study and to AI system A (if applicable).
They are given several task comprehension questions and are asked to complete an example task. \looseness=-1

In the second component, participants answer a total of eight questions (details in \cref{sec:studystimuli}).
They are told they can use any resources they want as in natural settings.
For each question, participants, except for those in the \cond{No-AI} condition, are provided with responses from AI system A (\cref{fig:introduction}).
The system's yes-or-no answers within their responses are correct for only half the questions.
In the \cond{Uncertain1st} and \cond{UncertainGeneral} conditions, the AI system expresses uncertainty in its answers for half the questions --- we refer to them as \textit{uncertain} answers versus \textit{not uncertain} answers.
We chose this breakdown to have sufficient data for each of the four possible scenarios of correct/incorrect answers with/without uncertainty expression.
We randomize the order in which questions are presented, as well as the set of questions for which the AI system expresses uncertainty.
However, since we based the AI system's answers on real responses from Copilot in Bing (see \cref{sec:studystimuli}), the set of answers and their correctness are fixed. \looseness=-1

In the final component, participants fill out an exit questionnaire about their experience with and perception of the AI system (if applicable), their background on LLMs, and basic demographic information; see Appendix D.
Lastly, participants are debriefed and reminded that some of the AI responses they saw may have contained inaccurate information. \looseness=-1

\subsection{Dependent Variables}
\label{sec:dvs}

We now formally define the dependent variables (DVs) that we measured, motivated in Section~\ref{sec:measuring}.
First, for each of the eight questions, we measured the following DVs based on participants' observed behavior:
\begin{itemize}[topsep=0pt]
    \item \var{Agree}: TRUE if the participant's final answer is the same as the AI system's answer; FALSE otherwise.
    \item \var{Correct}: TRUE if the participant's final answer is correct; FALSE otherwise.
    \item \var{Time}: Number of minutes from when the participant saw the task to when they clicked next.
    \item \var{LinkClick}: TRUE if the participant clicks on one or more links in the system's answer; FALSE otherwise. 
\end{itemize}

We additionally measured the following DVs based on participants' self-reported ratings:
\begin{itemize}[topsep=0pt]
    \item \var{UseAI}: TRUE if the participant selected ``AI system A's answer'' in the question ``What is your final answer based on? (Select all that apply)''; FALSE otherwise.
    \item \var{UseLink}: TRUE if they selected ``Your own reading of the linked sources in AI system A's answer'' in the above question; FALSE otherwise.
    \item \var{UseInternet}: TRUE if they selected ``Your own Internet search'' in the above question.; FALSE otherwise.
    \item \var{ConfidenceAI}: Rating on the question ``How confident are you in AI system A's answer?'' on a 5-point scale. 
    \item \var{ConfidenceAnswer}: Rating on the question ``How confident are you in your final answer?'' on a 5-point scale. 
\end{itemize}

Finally, based on responses to the exit questionnaire, we calculated the following indexes (all on a 5-point scale): 
\begin{itemize}[topsep=0pt]
    \item \var{TrustBelief}: Average rating on six statements adapted from the trust scale by McKnight et al.~\cite{McKnight2002}.
    \item \var{TrustIntention}: Average rating on four statements adapted from the trust scale by McKnight et al.~\cite{McKnight2002}.
    \item \var{Anthropomorphism}: Average rating on four items from the Godspeed Questionnaire Series~\cite{GSQ2009}.
    \item \var{Transparency}: Average rating on two statements: ``I feel I had a good understanding of what AI system A's answers were based on'' and ``I feel I had a good understanding of when AI system A's answers might be wrong.''
 \end{itemize}
    
Full details are in Appendix D.
Note that some DVs were not applicable for the \cond{No-AI} condition, where we measured only \var{Agree}, \var{Correct}, \var{Time}, \var{UseInternet}, and \var{ConfidenceAnswer}.
Here we made one (and only one) minor deviation from our pre-registration by including \var{Agree}.
Although participants in this condition do not see the AI system's answers, this gives us a baseline for how often participants would arrive at the same answer on their own. \looseness=-1

\subsection{Hypothesis \& Analysis}
\label{sec:analysis}

We expected the presence and perspective of uncertainty expression to impact participants' reliance, trust, and performance.
Formally, for each DV, we hypothesized that condition affects DV.
For each repeatedly measured DV, we additionally hypothesized that whether or not uncertainty was expressed in a particular AI response affects DV.
We tested our hypotheses with the following pre-registered, confirmatory analyses, for which we present results in \cref{sec:results}.

We first test the main effect of the conditions with a \textbf{between-condition analysis}. 
For repeatedly measured DVs, we fit the model \texttt{DV $\sim$ Condition + (1|participant) + (1|question)} with \cond{Control} as the reference level for \texttt{Condition}.
Then to compare the effects of the two conditions with uncertainty, we used a Wald test to test the equality of the corresponding coefficients. 
For DVs measured once in the exit questionnaire, we use analysis of variance (ANOVA) to compare means across the conditions.
If significant, we conduct pairwise comparisons with a post-hoc Tukey test.

Next, we test the effect of uncertainty being expressed or not in a particular response with a \textbf{within-condition analysis} for conditions with uncertainty expression.
For repeatedly measured DVs, we fit the model  \texttt{DV $\sim$ AIUncertain + (1|participant) + (1|question)}, where \var{AIUncertain} is TRUE if the AI response is uncertain and FALSE otherwise.
We fit this model once for data from the \cond{Uncertain1st} condition and once for data from \cond{UncertainGeneral}. \looseness=-1

We complement the confirmatory analyses with two additional analyses.
First, we conduct an exploratory analysis of the effect of AI's uncertainty expression on over- and underreliance by separately analyzing cases where the AI system gave correct versus incorrect answers.
Analysis details and results are presented in \cref{sec:separateanalysis}.
Second, we conduct a thematic analysis~\cite{boyatzis1998transforming,BraunClarke2006} of free-form responses from participants in the \cond{Uncertain1st} and \cond{UncertainGeneral} conditions describing their experience with and perception of the AI system. 
The first author drafted the codebook and conducted the initial coding, then discussed the results with all authors and refined the coding together.
We describe how the AI system's uncertainty expression affected participants in \cref{sec:results} along with the quantitative results, and describe how participants interpreted the system's uncertainty expression in \cref{sec:interpretation}.

\subsection{Questions and AI Responses Used}
\label{sec:studystimuli}

We selected a set of factual questions for participants to answer according to the following criteria: (1) most lay people should not know the answer; (2) the question and answer should not directly show up when using popular search engines; and (3) the answer can be objectively and automatically assessed.
To satisfy the criteria, we constructed a set of yes/no medical questions.
We began with questions from the MedQuAD dataset~\cite{MedQuAD2019} and made minor modifications to some to increase the difficulty of finding an answer.
We verified that each question does not show up as it is and can not be immediately answered using popular search engines, and consulted multiple sources to confirm the correct answer. 

To create AI responses that are realistic and reflect the state-of-the-art in LLM-infused search, we input the selected questions into Microsoft's Copilot in Bing.
All responses were obtained in July 2023. 
To keep the fluency, style, and content of responses as realistic as possible, we made only minor modifications such as presenting in-line citations using square brackets instead of superscripts (see \cref{fig:introduction}) and starting each response with ``Yes'' or ``No'' for consistency (most responses from Copilot in Bing did this already).
We did not make substantive changes to the content.

To create the uncertain versions of the responses, we drew on \citet{oba2022hedges}.
For \cond{Uncertain1st}, we replaced the leading ``Yes'' or ``No'' with an expression of the form ``\emph{I'm not \{certain, sure\}, but \{it seems to me, it seems like, I would guess, I'd guess that\}},''  selecting one phrase from each set of brackets.
Similarly, for \cond{UncertainGeneral}, we used an expression of the form ``\emph{\{It's unclear, It's not clear, There is uncertainty\}, but it seems like}.'' Current LLMs already output expressions such as ``\emph{I'm not sure, but}'' and ``\emph{I'm not sure, but my guess is}''~\cite{mielke-etal-2022-reducing}, so we believe these insertions preserve the realisticness of responses.
The only difference between the three versions of the AI system's responses is the presence and perspective of uncertainty expression; their information content is otherwise identical.

We selected the final eight questions such that:
(1) four questions have a correct answer of ``Yes'' and four have a correct answer of ``No'' so that always selecting ``Yes'' is no better or worse than random guessing; 
(2) the AI answers are correct for four and incorrect for four, so that always agreeing with the AI system is no better or worse than random guessing; 
(3) questions are not too easy to answer without access to the AI system (determined via piloting). 
The final set of questions used is provided in Appendix E along with the original responses from Copilot in Bing and our modified responses.
\looseness=-1


\begin{table*}[t!]
    \centering
    \caption[]{\textbf{Between-condition analysis.}
    \textnormal{We compare DVs across conditions. We report the model-estimated means (and standard errors) from our confirmatory analysis. The rightmost column shows pairs of conditions with statistically significant differences with significance marked as $^\ast$ ($p < 0.05$) or $^{\ast\ast}$ ($p < 0.01$).} \\ \\
    \footnotesize{\normalfont $^\dagger$Note that we did not compare all possible pairs of conditions (see \cref{sec:analysis}). For repeatedly measured DVs, we compared (\cond{Control} vs. \cond{Uncertain1st}/\cond{UncertainGeneral}/\cond{No-AI}) and (\cond{Uncertain1st} vs. \cond{UncertainGeneral}) but not (\cond{No-AI} vs. \cond{Uncertain1st}/\cond{UncertainGeneral}) to reduce the number of hypothesis testing. For DVs measured once in the exit questionnaire, we first compared the means of (\cond{Control}, \cond{Uncertain1st}, \cond{UncertainGeneral}) using ANOVA, then if significant, conducted pairwise comparisons.}
    }
    \resizebox{0.9\textwidth}{!}{
    \begin{tabular}{ccccccc}
    
        \hline
        
        Sec. & DV & \cond{Control} & \cond{Uncertain1st} & \cond{UncertainGeneral} & \cond{No-AI} & Significant differences$^\dagger$ \\
        
        \hline
        
        \multirow{2}{*}{\ref{sec:agree}} & \multirow{2}{*}{\var{Agree} (\%)} & \multirow{2}{*}{80.9\% (5.5)} & \multirow{2}{*}{74.8\% (6.7)} & \multirow{2}{*}{77.6\% (6.2)} & \multirow{2}{*}{58.4\% (8.5)} & \cond{No-AI} <$^{\ast\ast}$ \cond{Control} \\
          &  &  &  &  &  & \cond{Uncertain1st} <$^\ast$ \cond{Control} \\
          
        \hline
        \\ [-1.5ex]

        \multirow{3}{*}{\ref{sec:confidence}} & \var{ConfidenceAI} (1-5) & 3.95 (0.17) & 3.66 (0.17) & 3.80 (0.17) &  & \cond{Uncertain1st} <$^{\ast\ast}$ \cond{Control} \\ [2ex]
        
        & \var{ConfidenceAnswer} (1-5) & 4.30 (0.08) & 4.34 (0.08) & 4.27 (0.08) & 4.22 (0.08) &  \\ [1ex]
 
      \hline
      \\ [-1.5ex]

      \multirow{7}{*}{\ref{sec:source}} & \var{LinkClick} (\%) & 2.7\% (2.1) & 7.2\% (4.8) & 3.9\% (3.1) &  & \\ [2ex]
      
      & \var{UseAI} (\%) & 77.3\% (5.1) & 64.8\% (6.9) & 72.3\% (6.0) &  & \\ [2ex]
      
      & \var{UseLink} (\%) & 74.7\% (5.5) & 85.5\% (4.0) & 81.7\% (4.8) &  & \\ [2ex]
      
      & \var{UseInternet} (\%) & 19.1\% (5.6) & 27.0\% (7.4) & 23.2\% (6.7) & 92.9\% (2.5) & \cond{Control} <$^{\ast\ast}$ \cond{No-AI} \\ [1ex]
        
      \hline
      \\ [-1.5ex]
      
      \multirow{6}{*}{\ref{sec:trust}} & \var{TrustBelief} (1-5) & 3.90 (0.06) & 3.86 (0.07) & 4.00 (0.07) &  & \\ [1ex]
      
      & \multirow{2}{*}{\var{TrustIntention} (1-5)} & \multirow{2}{*}{3.25 (0.10)} & \multirow{2}{*}{2.91 (0.10)} & \multirow{2}{*}{3.36 (0.10)} & \multirow{2}{*}{} & \cond{Uncertain1st} <$^{\ast}$ \cond{Control} \\
      &  &  &  &  &  & \,\,\,\,\,\,\,\,\, <$^{\ast}$ \cond{UncertainGeneral} \\ [1ex]
      
      & \var{Anthropomorphism} (1-5) & 3.07 (0.10) & 3.00 (0.11) & 3.13 (0.11) &  & \\ [2ex]
      
      & \var{Transparency} (1-5) & 4.04 (0.06) & 3.93 (0.07) & 4.01 (0.07) &  & \\ [1ex]

      \hline
      \\ [-2ex]
      
      \multirow{3}{*}{\ref{sec:correct}} & \multirow{2}{*}{\var{Correct} (\%)} & \multirow{2}{*}{63.9\% (8.6)} & \multirow{2}{*}{72.8\% (7.4)} & \multirow{2}{*}{67.9\% (8.1)} & \multirow{2}{*}{74.2\% (7.1)} & \cond{Control} <$^{\ast\ast}$ \cond{Uncertain1st} \\
      &  &  &  &  &  & \cond{Control} <$^{\ast\ast}$ \cond{No-AI} \\ [1ex]

       & \var{Time} (min) & 2.13 (0.22) & 2.10 (0.23) & 2.03 (0.22) & 1.57 (0.21) & \cond{No-AI} <$^\ast$ \cond{Control} \\ [1ex]
       
       \hline
       
    \end{tabular}
    }
    \label{tab:analysis1}
\end{table*}

\subsection{Data Collection and Participants}
\label{sec:datacollection}

We conducted our experiment on Amazon Mechanical Turk (MTurk), a crowdsourcing platform widely used for human-subject experiments.
Research has shown data from MTurk workers is comparable to data from other pools (e.g., commercial panels, social media, colleges)~\cite{Berinsky2012MTurk,Buhrmester2011MTurk,Casler2013MTurk,coppock2019MTurk,Mason2012MTurk,Paolacci2014MTurk}, but recently there has been a decrease in data quality~\cite{veselovsky2023mturk,Chmielewski2020MTurk,Webb2022MTurk,Marshall2023MTurk,kennedy2020MTurk,lu2022MTurk}.
Indeed, in pilot studies we found that a strong requirement on qualification was necessary to obtain meaningful data, with the highest quality data obtained when requiring a ``Masters'' qualification (granted by Amazon based on past performance).
Since the pool of available participants with a Masters qualification is limited, we pre-registered a recruitment plan in which we would initially aim to recruit 432 (determined via a power analysis) U.S.-based MTurk workers with a Masters qualification, 99\% or higher approval rating, and at least 2000 completed human intelligence tasks (HITs), removing the Masters requirement after 7 days if we were unable to meet our target sample size.
In parallel, we implemented best practices to mitigate the use of bots and improve data quality, including CAPTCHAs, honeypot questions, speed checks, attention checks, and open-ended questions, which we used to define data exclusions. \looseness=-1

We collected 656 complete responses over the course of two weeks in September 2023, of which we excluded 252 (38.4\%) based on five pre-registered exclusion criteria.
Our final sample consists of 404 responses: 104 in \cond{Control}, 92 in \cond{Uncertain1st}, 94 in \cond{UncertainGeneral}, and 114 in \cond{No-AI}.
See Appendix B for more on our data collection procedures. \looseness=-1

Participants were paid \$5 USD.
The payment was determined based on the expected experiment duration of 20 minutes (estimated from pilot studies) and target hourly wage of \$15.
The actual median experiment duration was 20.5 minutes, so on average, participants were paid \$14.80 per hour.
See Appendix A for more information about participants.

\section{Results: Confirmatory analysis}
\label{sec:results}

We now present the results of our pre-registered, confirmatory analyses.
\cref{tab:analysis1,tab:analysis2} contain the between-condition and within-condition analysis results, respectively.
We refer the reader to Section~\ref{sec:dvs} for definitions of all DVs.
We present the estimated means (and standard errors) from the fitted models, calculated without conditioning on the random effects, and represent binary variables in percentages.
We use \textit{significance} to refer to statistical significance at the level of $p < 0.05$. \looseness=-1

\subsection{Agreement with AI: \var{Agree}}
\label{sec:agree}

We begin with the results of agreement, a commonly used behavioral measure of reliance and trust.
Our first finding is that \textbf{people tend to agree with the AI system when its responses are provided}. 
Our between-condition analysis suggests that participants with access to the AI system are significantly more likely to submit the same answer as the system than those who do not have access (80.9\% \cond{Control} vs. 58.4\% \cond{No-AI}).
We next find that \textbf{AI's uncertainty expression decreases agreement with the AI system}.
Compared to \cond{Control} (80.9\%), \var{Agree} is significantly lower in \cond{Uncertain1st} (74.8\%).
It is also lower in \cond{UncertainGeneral} (77.6\%), although the difference is not significant. 
Our within-condition analysis suggests AI's uncertainty expression decreases agreement at the instance level as well.
The estimated means of \var{Agree} for not uncertain vs. uncertain AI responses are 84.7\% vs. 70.9\% for \cond{UncertainGeneral} (significantly different) and 79.5\% vs. 73.4\% for \cond{Uncertain1st} (not significantly different).
We find support for this finding in the qualitative data as well. 
10 participants (out of 186 in \cond{Uncertain1st} and \cond{UncertainGeneral}) stated that when they disagreed with the system it was, as one put it, ``\emph{because of the uncertainty of the answers provided by AI}.''

Other reasons participants mentioned for disagreement included the system's answer being different from their own knowledge (e.g., ``\emph{A couple of the AI's answers didn't make sense so from my own common sense I had to make my own judgement}'') or the information in other resources (e.g., ``\emph{The answer seemed to contradict the links given, or I could not find how they came to that answer with the information from the links given}'') and having lower trust in the system. \looseness=-1

\subsection{Confidence in Answers: \var{ConfidenceAI}, \var{ConfidenceAnswer}}
\label{sec:confidence}

Looking at participants' self-reported confidence in answers, we find that \textbf{AI's uncertainty expression decreases people's confidence in its answer}.
Compared to \cond{Control} (3.95 on a 5-point scale), \var{ConfidenceAI} is significantly lower in \cond{Uncertain1st} (3.66). It is also lower in \cond{UncertainGeneral} (3.80), although the difference is not significant.
Our within-condition analysis provides further evidence for this finding and suggests that \textbf{AI's uncertainty expression decreases people' confidence in their final answer at the instance level}.
For both \cond{Uncertain1st} and \cond{UncertainGeneral}, \var{ConfidenceAI} and \var{ConfidenceAnswer} are significantly lower on instances with uncertain (vs. not uncertain) AI responses.
Indeed, one participant in the \cond{Uncertain1st} condition stated, ``\emph{If the AI didn't seem confident, I would like [sic] on the links. If the AI seemed confident I assumed he was correct most of the time}.''


\begin{table*}[t!]
    \centering
    \caption{
    \textbf{Within-condition analysis.}
    \textnormal{For each of the two conditions with uncertainty, we compare DVs measured on instances with AI answers that are not uncertain vs. uncertain. We report the model-estimated means (and standard errors) from our confirmatory analysis. > and < note statistically significant differences with significance marked as $^\ast$ ($p < 0.05$) or $^{\ast\ast}$ ($p < 0.01$).} \\ \\
    \footnotesize{\normalfont $^\dag$The model does not fit properly due to large individual variance which we discuss in an exploratory analysis in Appendix C.1.}
    }
    \resizebox{0.85\textwidth}{!}{
    \begin{tabular}{c@{\hskip 0.6cm}c@{\hskip 0.6cm}clc@{\hskip 1cm}clc}
        \hline
         \multirow{2}{*}{Sec.} & \multirow{2}{*}{DV} & \multicolumn{3}{c}{\cond{Uncertain1st}} & \multicolumn{3}{c}{\cond{UncertainGeneral}} \\
          & & Not Uncertain & & Uncertain & Not Uncertain & & Uncertain \\
         \hline
         
         \ref{sec:agree} & \var{Agree} (\%) & 79.5\% (7.2) & & 73.4\% (8.6) & 84.7\% (5.0) & >$^{\ast\ast}$ & 70.9\% (7.8) \\
         
         \hline
         
         \multirow{2}{*}{\ref{sec:confidence}} & \var{ConfidenceAI} (1-5) & 3.88 (0.20) & >$^{\ast\ast}$ & 3.44 (0.20) & 3.95 (0.16) & >$^{\ast\ast}$ & 3.65 (0.16) \\
         
         & \var{ConfidenceAnswer} (1-5) & 4.42 (0.09) & >$^{\ast\ast}$ & 4.26 (0.09) & 4.36 (0.07) & >$^{\ast\ast}$ & 4.18 (0.07) \\
     
         \hline
         
         \multirow{4}{*}{\ref{sec:source}} & \var{LinkClick} (\%) & 14.3\% (6.9) &  & 9.9\% (5.1) & \multicolumn{3}{c}{\emph{\small See table caption}$^\dag$}  \\
    
          & \var{UseAI} (\%) & 73.2\% (7.2) & >$^{\ast\ast}$ & 57.4\% (8.9) & 79.7\% (4.7) & >$^{\ast\ast}$ & 62.9\% (6.5) \\
          
          & \var{UseLink} (\%) & 86.5\% (4.2) &  & 84.3\% (4.7) & \multicolumn{3}{c}{\emph{\small See table caption}$^\dag$} \\
          
          & \var{UseInternet} (\%) & 23.0\% (6.3) & <$^{\ast\ast}$ & 34.1\% (7.8) & 19.2\% (6.5) &  & 26.9\% (8.1) \\
         
         \hline
         
         \multirow{2}{*}{\ref{sec:correct}} & \var{Correct} (\%) & 73.6\% (8.9) &  & 75.7\% (8.4) & 69.4\% (10.1) &  & 70.3\% (9.9) \\
 
         & \var{Time} (min) & 2.00 (0.29) &  & 2.19 (0.29) & 1.84 (0.25) & <$^\ast$ & 2.23 (0.25) \\
         
         \hline
        
    \end{tabular}
    }
    \label{tab:analysis2}
\end{table*}

\subsection{Source Usage: \var{LinkClick}, \var{UseAI}, \var{UseLink}, \var{UseInternet}}
\label{sec:source}

Looking at source usage, we find that \textbf{people with access to the AI system conduct their own Internet search less frequently than those without access}.
\var{UseInternet} is notably lower in \cond{Control} (19.1\%) than \cond{No-AI} (92.9\%). 
However, we see no significant differences in source usage between the three conditions in which AI responses are present, meaning \textbf{there is no evidence that the presence and perspective of AI's uncertainty expression affect people's source usage behavior}.
On the other hand, from our within-condition analysis, we find that AI's uncertainty expression has a significant effect on the self-reported DVs at the instance level.
On instances with uncertain AI responses (vs. not uncertain responses), \var{UseAI} is significantly lower in both \cond{Uncertain1st} and \cond{UncertainGeneral}, and \var{UseInternet} is significantly higher in \cond{Uncertain1st}.
These results suggest that \textbf{at the instance level, AI's uncertainty expression decreases the use of the system's answer and increases the use of other resources.}
In their free-form responses, 11 participants stated that the system's uncertainty motivated them to verify information using the links provided in the answer (e.g., ``\emph{I clicked on the links provided when the AI was uncertain of the answers}'') while 13 mentioned it motivated them to perform their own search (e.g., ``\emph{When the AI wasn't certain, I searched on Google}''). \looseness=-1

We note that both \var{LinkClick} and \var{UseLink} are intended to capture whether or not participants read the linked sources provided in the AI system's responses, but the estimated means of \var{LinkClick} (2.7\% -- 7.2\%) from the between-condition analysis are much lower than those of \var{UseLink} (74.7\% --- 85.5\%).
There are several factors that might contribute to this discrepancy.
First, these estimated means are from different models that include participants and questions as random effects.
The intercepts for random effects in these models are quite high.
Looking at the raw data, there is still a gap, but not as large: the actual means are 30.6\% -- 34.9\% for \var{LinkClick} and 64.2\% -- 71.6\% for \var{UseLink}.
Second, there could have been measurement error from self-report bias for \var{UseLink}~\cite{donaldson2002selfreportbias}.
Finally, some participants could have considered their answers to be based on their ``own reading of the linked sources'' if they read the list of links, even if they didn't click to open them.
For example, one participant wrote, ``\emph{If the link title gave the same answer as the AI answer, then I assumed it was the right answer.}''
This discrepancy emphasizes the value of including both behavioral and self-reported measures.

\subsection{Trust and Perception of AI: \var{TrustBelief}, \var{TrustIntention}, \var{Anthropomorphism}, \var{Transparency}}
\label{sec:trust}

Moving onto trust and perception of AI, we find that \textbf{the presence and perspective of uncertainty expression neither affect people's trust beliefs nor the perceived anthropomorphism and transparency of the system.}
There are no significant differences in \var{TrustBelief}, \var{Anthropomorphism}, and \var{Transparency} between \cond{Control} and the two conditions with uncertainty.
Overall, participants had \textbf{somewhat positive trust beliefs} about the system (\var{TrustBelief} is around ``4: Somewhat agree'' for all conditions); reported that they had a \textbf{somewhat good understanding of what the AI system's answers were based on and when they might be wrong} (\var{Transparency} is around ``4: Somewhat agree''); and had \textbf{neutral perceptions of anthropomorphism} (\var{Anthropomorphism} is around ``3: Neutral'').  \looseness=-1

In contrast, we find that \textbf{first-person expressions of uncertainty decrease trust intentions while expressions from a general perspective do not}.
\var{TrustIntention} is significantly lower in \cond{Uncertain1st} (2.91) compared to both \cond{Control} (3.25) and \cond{UncertainGeneral} (3.36), indicating a lower desire to use the system. 
Illustrative of this, one participant in the \cond{Uncertain1st} condition stated, ``\emph{[The AI system] was very non-committal in its answers so I didn't feel I could trust it}.''
This suggests that frequent first-person expressions of uncertainty can lead people to view the system as less trustworthy and decrease their desire to use it.
More about the used scales and item-level results is in Appendix C.2. \looseness=-1

\subsection{Task Performance: \var{Correct}, \var{Time}}
\label{sec:correct}\label{sec:time}

Finally, we analyze participants' task performance.
From our between-condition analysis of \var{Correct}, we first find that \textbf{having access to the AI system decreases people's accuracy}.
Participants with access to the system have significantly lower accuracy than those without (63.9\% \cond{Control} vs. 74.2\% \cond{No-AI}).
However, this result should be interpreted in the context of the AI system's low overall accuracy (50.0\% in our experimental setup).
Second, we find that \textbf{AI's uncertainty expression increases people's accuracy}.
\var{Correct} is significantly higher in \cond{Uncertain1st} (72.8\%) than \cond{Control} (63.9\%).
It is also higher in \cond{UncertainGeneral} (67.9\%), but the difference is not significant. \looseness = -1

From our between-condition analysis of \var{Time}, we find that \textbf{having access to the AI system increases task time}. \var{Time} is significantly higher in \cond{Control} (2.13 min) than \cond{No-AI} (1.57 min).
Together with the results on correctness, this suggests that \textbf{having access to the AI system decreases overall performance in our experimental setup}.
Between \cond{Control} and the two conditions with uncertainty, there are no significant differences in task time.
However, our within-condition analysis suggests that \textbf{AI's uncertainty expression increases task time at the instance level}.
The estimated means of \var{Time} for not uncertain vs. uncertain AI responses are 1.84 min vs. 2.23 min for \cond{UncertainGeneral} (significantly different) and 2.00 min vs. 2.19 min for \cond{Uncertain1st} (not significantly different). 
An explanation of this result is that AI's uncertainty expression slows people down to use more caution when completing the task.

\begin{figure*}[t!]
\centering
\begin{subfigure}[t]{0.49\textwidth}
    \centering
    \includegraphics[width=\columnwidth]{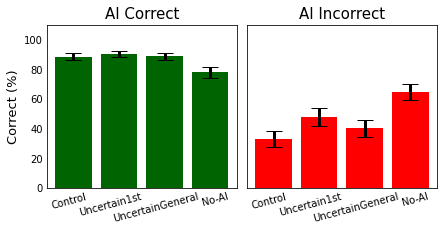}
    \caption{Between-condition analysis results.}
    \label{fig:appropriatereliance_h1}
\end{subfigure}
\begin{subfigure}[t]{0.49\textwidth}
    \centering
    \includegraphics[width=\columnwidth]{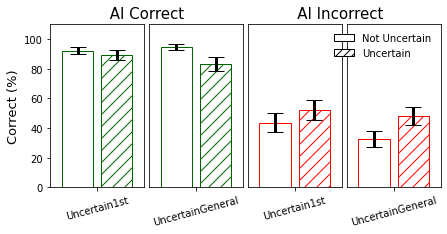}
    \caption{Within-condition analysis results.}
    \label{fig:appropriatereliance_h2}
\end{subfigure}
\caption{\textbf{Exploratory analysis of over- and underreliance (\cref{sec:separateanalysis}).} \textnormal{We analyze \var{Correct} (\%) separately for questions the system answered correctly vs. incorrectly. We show the model-estimated means and standard errors for each condition (\cref{fig:appropriatereliance_h1}) and for not uncertain vs. uncertain responses in the conditions with uncertainty (\cref{fig:appropriatereliance_h2}).}}
\Description{Figure 2a consists of two bar plots. The first bar plot for "AI Correct" instances show 88.5\% for Control, 90.2\% for Uncertain1st, 88.7\% for UncertainGeneral, and 77.9\% for No-AI. The second bar plot for "AI Incorrect" instances show 33.0\% for Control, 47.8\% for Uncertain1st, 40.1\% for UncertainGeneral, and 64.7\% for No-AI.
Figure 2b consists of bar plots comparing results for not uncertain vs. uncertain responses in the conditions with uncertainty. For "AI Correct" instances, the plots show 92.9\% vs. 89.2\% for Uncertain1st and 94.8\% vs. 83.1\% for UncertainGeneral. For "AI Incorrect" instances, the plots show 43.6\% vs. 52.0\% for Uncertain1st and 32.8\% vs. 48.0\% for UncertainGeneral.}
\label{fig:appropriatereliance}
\end{figure*}

\section{Results: Additional analyses}
\label{sec:exploratory}

\subsection{Effect of Uncertainty Expression on Over- and Underreliance}
\label{sec:separateanalysis}

In Section~\ref{sec:agree}, we analyzed participants' agreement with the AI system as a measure of reliance.
Agreement can be appropriate or inappropriate, depending on the correctness of the AI system's answers; agreeing with the system when it is incorrect is a sign of overreliance, whereas disagreeing with it when it is correct is a sign of underreliance.
To better understand the extent to which uncertainty expression leads to either of these phenomena, we take inspiration from the analysis of ~\citet{chen2023understanding} and separately analyze participants' agreement on questions the AI system answers correctly and questions it answers incorrectly.
Note that, having conditioned on the (in)correctness of the AI system, analyzing agreement is equivalent to analyzing correctness of people's final answers.
We present the results here in terms of correctness to focus attention on whether reliance is appropriate (i.e., beneficial to the user) or not. 

Concretely, we run similar analyses to those presented in Section~\ref{sec:correct}, but fit the regression models once on data from the four questions the system answered correctly and once on data from the four questions the system answered incorrectly (see \cref{sec:studystimuli}).
We show the estimated means (and standard errors) from these models in \cref{fig:appropriatereliance}. \looseness=-1

First, by comparing the \cond{Control} and \cond{No-AI} conditions, we observe that having access to the AI system's answer increases accuracy when the system is correct (the estimated mean of \var{Correct} is 88.5\% in \cond{Control} vs. 77.9\% in \cond{No-AI}), but decreases accuracy when it is incorrect (33.0\% in \cond{Control} vs. 64.7\% in \cond{No-AI}), as in \cref{fig:appropriatereliance_h1}.
Comparing \cond{Control} with \cond{Uncertain1st} and \cond{UncertainGeneral}, we see that having the AI system express uncertainty improves accuracy on questions that the system answers incorrectly without reducing accuracy when the system is correct.
In line with our earlier results, expressing uncertainty in the first-person perspective leads to a bigger improvement in accuracy when the AI system is incorrect compared with expressing uncertainty in the general perspective.

To better understand how expressions of uncertainty drive accuracy, we break down the results further, comparing task accuracy on questions for which the system expresses uncertainty and those for which it does not (\cref{fig:appropriatereliance_h2}).
We find that expressing uncertainty about a particular question leads to some reduction in accuracy when the AI system is correct (92.2\% to 89.2\% for \cond{Uncertain1st}, 94.8\% to 83.1\% for \cond{UncertainGeneral}), but a greater increase in accuracy when the AI system is incorrect (43.6\% to 52.0\% for \cond{Uncertain1st}, 32.8\% to 48.0\% for \cond{UncertainGeneral}).

While these results provide some evidence that expressions of uncertainty help reduce overreliance, we note that participants in the \cond{Uncertain1st} and \cond{UncertainGeneral} conditions still have substantially lower accuracy on questions where the AI system is incorrect compared with participants in the \cond{No-AI} condition.

\subsection{Participants' Interpretations of AI's Uncertainty Expression}
\label{sec:interpretation}

While our quantitative results shed light on whether expressions of uncertainty impact reliance and trust, they cannot tell us why.
We next explore participants' interpretations of the expressed uncertainty via a thematic analysis of free-form responses to the question ``When and why do you think AI system A expresses uncertainty?'' in the exit questionnaire. \looseness = -1

The majority of participants (N=102 of the 186 in the conditions with uncertainty) attributed the system's expressed uncertainty to its inability to answer a particular question, for example because it could not find an answer, found conflicting or unreliable answers, or could not understand the information it found. 
Three suggested the system was programmed to express uncertainty, as in ``\emph{I would guess there's some sort of certainty variable and if the score is below a level, an uncertainty message is included in the result}.''
Another ten suggested the expressed uncertainty is due to the inherent difficulty of the question. As one put it, ``\emph{It could be a question that is very hard to come up with a simple yes or no answer}.'' \looseness=-1

These interpretations are all in line with the goal of reducing overreliance: if the system is unable to answer a question or the question is inherently difficult, users should verify the answer for themselves. 
Five participants explicitly interpreted the expressed uncertainty as a way of encouraging users to check their answers, for example, ``\emph{I suppose the AI wanted us to do further research in those cases where it could not be 100\% sure of the answer}.''

A small number of participants attributed the uncertainty to other reasons, such as impression management (``\emph{to appear more human, encourage confidence, and appear thoughtful}''), maintaining credibility (``\emph{It doesn't want to risk being wrong on something and ruining it's credibility}''), avoiding liability (``\emph{it expresses uncertainty to absolve it of responsibility in the event it is wrong}''), or restrictions on answering medical questions (``\emph{the AI is programmed not to dispense medical advice which could potentially be harmful without a caveat}'').
These interpretations may not necessarily reduce overreliance. \looseness=-1

One might ask whether participants interpreted the two types of uncertainty expression differently.
Indeed, we found that participants in the \cond{UncertainGeneral} condition were more likely than those in \cond{Uncertain1st} (51.5\% vs. 41.3\%) to attribute the uncertainty to the AI system finding conflicting or unreliable information or the question being inherently hard, whereas those in \cond{Uncertain1st} were more likely (20.7\% vs. 7.4\%) to attribute it to limitations of the AI system itself. \looseness=-1

\section{Discussion}
\label{sec:discussion}

Our results suggest that expressing uncertainty through natural language can be an effective way to reduce overreliance and (over)trust in LLM-infused search engines.
Expressions of uncertainty led to more cautious behaviors, from taking longer to arrive at an answer to reporting more reliance on outside sources.
However, it did not fully eliminate overreliance; the participants with the highest task performance were those without access to AI responses.

We find that perspective matters: uncertainty expressions in first-person show stronger effects than general perspective. 
This is consistent with prior findings that first-person messages increase recipients' involvement and engagement~\cite{chen2017persuasive,Zhang2022CHI} compared with general or third-person messages.
This persuasive effect of first-person expressions should be interpreted with caution: while it helps heighten the warning effect of a negative message such as uncertainty, it might amplify a positive message, even if unjustified. 
For example, one may want to avoid first-person expressions of \emph{confidence} because they may exacerbate overreliance and over-trust, as found in prior work~\cite{zhou2024relying}.
There are also concerns around harms from anthropomorphism of AI systems that may stem from over-trust, deception, threats to human agency, and propagation of stereotypes~\cite{abercrombie2023mirages}.
While we did not observe that first-person uncertainty expression increases perceived anthropomorphism, people can start assigning social attributes to machines without conscious awareness~\cite{nass2000machines}.
Future research and practices should further explore the long-term effects of interacting with AI systems expressing uncertainty in a first-person perspective and consider other potential negative effects of anthropomorphism. \looseness=-1

Our research has implications both for those building and deploying LLMs and LLM-infused applications and for policymakers regulating the use of AI.
Most critically, any approach to reducing overreliance should be validated through empirical research.
There may also be tradeoffs when balancing over- and underreliance.
In our study, the most successful approach to reducing overreliance was to use first-person uncertainty expression, but this also decreased participant trust in the AI system, which may be undesirable in settings where people already under-trust the AI system.
We believe there is no one-size-fits-all approach to implementing natural language uncertainty expression. 
For these reasons, given that the issue of how to manage overreliance is of particular importance to regulators, we advocate for raising awareness of the complexities of mitigating overreliance and for customized, evidence-based solutions, rather that universal ones.
\looseness=-1

There are limitations to our research.
The widespread deployment of LLMs is still relatively new and the human-computer interaction and broader research communities are still grappling with the question of how to design effective studies to understand how end users perceive and interact with them. There is always a tradeoff between the controllability of the experiment and the generalizability of the conclusions to user behaviors in their day-to-day tasks~\cite{mcgrath1995methodology}.
In order to be able to measure agreement and correctness, we adopted an experimental set-up inspired by the AI-assisted decision-making literature in which study participants provide simple yes/no answers to questions.
This approach does not allow exploring how the expression of uncertainty would impact people's behavior when completing more complex tasks, like writing an article or planning a trip.
Further, our measurements of time and source usage are less reliable than they would have been had we opted for an in-person lab study.
We chose questions from the medical domain, where overreliance is particularly concerning, but people may behave differently when seeking information about their own medical symptoms rather than answering a pre-defined set of questions.
People may also behave differently when given the chance to interact with the system repeatedly as opposed to in a single session.
The AI system in our study exhibited low accuracy and expressed uncertainty often, in a poorly calibrated manner.
These design choices may have impacted our results --- particularly the lower task performance when given access to the AI system.
Also, there may be differences across cultures and languages in how people interpret or react to uncertainty.
Our study was conducted in English with U.S.-based participants and results may not generalize to other cultural and linguistic contexts.

For all of these reasons, while our findings suggest that natural language expressions of uncertainty could be an effective approach to reducing overreliance, we caution against overgeneralizing from our study.
Instead, we view our results as evidence that language choices matter in how people perceive and act on the outputs of LLMs, and teams building and deploying LLMs should therefore evaluate them carefully with end users before release.

\section{Ethical Considerations and Positionality}

We conclude with a reflection on the ethical considerations of our work and our positionality.

\ \\ \noindent \textbf{Mitigating harms to human subjects.}
We recruited U.S.-based participants on MTurk, which many people rely on as a primary source of income. As discussed in Section~\ref{sec:datacollection}, we aimed to provide an hourly wage of \$15 USD. We came close to this goal, with participants receiving an estimated \$14.80 per hour on average. (This is likely an underestimate of average wage, since we have no way to know if workers spent time on other activities between accepting the task and completing it.) This is substantially higher than the U.S. federal minimum wage of \$7.25 per hour, though a few states have recently adopted a minimum wage of \$15/hour or higher. As discussed in the FAccT 2023 panel ``The Humans Behind the Intelligence: Speaking with Data Workers,'' our choice to limit participation to workers with a 99\% or higher approval rating, at least 2000 completed tasks, and in some cases, a Masters qualification, prevented workers who are new to MTurk from participating. We made the decision to include these qualifications after piloting several versions of the study with less restrictive qualifications and finding that the data quality was too poor to use. We paid and approved the work of everyone who completed the study, regardless of whether their responses passed our quality checks. At the end of the study, we debriefed participants, reminding them that the medical information output by the AI system was sometimes incorrect. We did not collect personally identifiable information except for MTurk IDs, which were used to ensure that workers who participated in pilots of our study did not participate in the main study. These were deleted when no longer needed. Our procedure was reviewed and approved by our internal IRB and we obtained participant consent. \looseness=-1

\ \\ \noindent \textbf{Potential negative societal impact.}
While our results provide evidence for the effectiveness of natural language expressions of uncertainty for reducing overreliance, generalizing too heavily from our findings could lead to potential harms.
Given the limitations of our research (see \cref{sec:discussion}), teams deploying LLMs or LLM-infused applications should not make decisions about how to express uncertainty to end users without extensive user testing in their own contexts.
They also should not assume that they have addressed overreliance by expressing uncertainty. (Indeed, in our study we see that participants still have higher task performance with no access to the AI system.) Likewise, regulators should avoid making blanket requirements on uncertainty expression, at least until more research has been done.

Separate from these potential unintentional misuses of our research, there is a possibility that bad actors could strategically incorporate the expression of uncertainty into an LLM's output to make them more persuasive, regardless of whether they represent objective fact, potentially contributing to the spread of misinformation. \looseness=-1

\ \\ \noindent \textbf{Positionality.}
Our research questions and design were influenced by our position as employees of a U.S.-based technology company. Members of our research team have first-hand experience observing and participating in discussions regarding the responsible development and deployment of LLM-infused applications, which has shaped our understanding of gaps in knowledge and other practical challenges that arise in attempting to meet responsible AI principles and proposed regulatory requirements. We had access to sufficient budget to run large-scale experiments, which is not an option for some research teams. Our view that there are potential benefits of responsibly deployed LLM-infused applications is likely influenced by our experience in industry, yet we acknowledge some fundamental limitations of using LLMs for information retrieval~\cite{Shah2022}. We encourage future research on overreliance from research teams in academia and civil society. \looseness=-1

\begin{acks}
We thank the participants for their time and effort and the anonymous reviewers for helpful comments.
We also thank Jake Hofman, Dan Goldstein, Sophie Spatharioti, Amy Heger, Samir Passi, Harsh Kumar, Wesley Deng, Snehal Prabhudesai, Abhraneel Sarma, Yeji Park, Nina Grgić-Hlača, Julia Stoyanovich, Jessica Hullman, and the Microsoft Research New York City lab and FATE group for thoughtful discussion and feedback.
SK acknowledges support from the NSF Graduate Research Fellowship and the Princeton SEAS Howard B. Wentz, Jr. Junior Faculty Award to her PhD advisor.
\end{acks}

\bibliographystyle{ACM-Reference-Format}
\bibliography{main}

\appendix

\renewcommand\thefigure{\thesection\arabic{figure}}
\renewcommand\thetable{\thesection\arabic{table}}

\setcounter{footnote}{2}

\section*{Appendix}

The appendix is structured in the following way.
\begin{itemize}[leftmargin=*]
    \item \textbf{\cref{appendix:participants}:} Participant Demographics and Background
    \item \textbf{\cref{appendix:datacollection}:} Data Collection and Exclusion
    \item \textbf{\cref{appendix:results}:} Additional Analyses
    \begin{itemize}
        \item \textbf{\cref{appendix:link}:} Exploration of \var{LinkClick} and \var{UseLink}
        \item \textbf{\cref{appendix:trust}:} Internal Consistency of Scales Used to Measure Trust and Perception of AI
    \end{itemize}
    \item \textbf{\cref{appendix:fullexperiment}:} Full Wording Used in the Experiment
    \item \textbf{\cref{appendix:studystimuli}:} Questions and AI Responses Used
\end{itemize}

\section{Participant Demographics and Background}
\label{appendix:participants}

In the exit questionnaire, participants optionally self-reported their gender, age, race, ethnicity, and level of education. Of 404 participants in the final sample, 51.7\% identified as woman, 46.8\% as man, and 0.5\% as non-binary. The age distribution was as follows: 18--24 (5.2\%), 25--34 (32.4\%), 35--44 (35.1\%), 45--54 (14.4\%), 55-64 (9.7\%), 65--74 (2.7\%), and 75--84 (0.2\%). Regarding race, 82.4\% identified as white alone, followed by Asian alone (6.7\%), Black or African American alone (5.9\%), two or more races (2.0\%), and American Indian and Alaska Native alone (1.0\%).  Regarding ethnicity, 12.4\% identified as Hispanic or Latino, and 86.9\% as not. For the highest level of education completed, 58.7\% said 4 year degree, followed by professional degree (13.6\%), high school graduate (9.4\%), some college (7.9\%), 2 year degree (7.7\%), doctorate (1.7\%), and less than high school (0.2\%).

In \cref{tab:participantinfo}, we compare the demographics of our study sample (N=404) with U.S. Census data~\cite{us2020census,census2022education}. Compared to the U.S. population, our sample is younger and more educated, a common characteristic of samples recruited through online crowdsourcing platforms~\cite{Huff2015MTurk,paolacci2010MTurk,Ross2010MTurk,Hitlin2016MTurk}. Further, white respondents are over-represented, while Black and Hispanic/Latino respondents are under-represented.

In the exit questionnaire, we also measured the following variables based on participants' self-reported ratings on a 5-point scale:
\begin{itemize}[leftmargin=*,topsep=0pt]
    \item \var{LLMFamiliarity}: Rating on the question ``How familiar are you with LLMs and LLM-infused applications such as OpenAI’s ChatGPT, Microsoft’s Bing Chat, and Google’s Bard?'' 
    \item \var{LLMUse}: Rating on the question ``How often do you use LLMs and LLM-infused applications such as OpenAI’s ChatGPT, Microsoft’s Bing Chat, and Google’s Bard?''
    \item \var{LLMAttitude}: Rating on the question ``Overall, how do you feel about LLMs and LLM-infused applications such as OpenAI’s ChatGPT, Microsoft’s Bing Chat, and Google’s Bard?'' 
\end{itemize}

The mean and standard deviation of \var{LLMFamiliarity} are $3.0 \pm 1.0$ around ``3: Moderately familiar, I know what they are and can explain.''
For \var{LLMUse}, they are $2.8 \pm 1.2$ between ``2: Rarely, about 1--2 times a month'' and ``3: Sometimes, about 3--4 times a month.''
For \var{LLMAttidue}, they are $3.8 \pm 1.0$ between ``3: Neutral'' and ``4: Somewhat positive.''
This result consistent with that of \citet{koonchanok2023tracking} who investigated public attitudes towards ChatGPT by analyzing Twitter data from December 2022 to March 2023 and found that the overall sentiment is largely neutral to positive.

\begin{table*}[ht!]
\caption{\textbf{Demographics of our study sample (N=404) compared to the U.S. population.} \textnormal{N/A indicates there is no data for the given category. $^\ast$ indicates multiple selection and self-description were possible.}}
\centering
\begin{tabular}{llrr}
    \hline
    Demographic Attribute & & Sample & Census \\
    \hline
    \multirow{6}{*}{Gender$^\ast$} & Woman & 51.7\% & N/A \\
           & Man & 46.8\% & N/A \\
           & Non-binary & 0.5\% & N/A \\
           & Gender-diverse & 0.0\% & N/A \\
           & Transgender & 0.0\% & N/A \\
           & Prefer not to answer & 1.0\% & N/A \\
    \hline
    \multirow{2}{*}{Sex} & Female & N/A & 50.9\% \\
        & Male   & N/A & 49.1\% \\
    \hline
    \multirow{9}{*}{Age} & 18--24 & 5.2\% & 9.4\% \\
        & 25--34 & 32.4\% & 13.5\% \\
        & 35--44 & 35.1\% & 12.7\% \\
        & 45--54 & 14.4\% & 12.4\% \\
        & 55--64 & 9.7\% & 13.1\% \\
        & 65--74 & 2.7\% & 10.0\% \\
        & 75--84 & 0.2\% & 4.9\% \\
        & 85 or older & 0.0\% & 1.9\% \\
        & Prefer not to answer & 0.2\% &  N/A\\
    \hline
    \multirow{8}{*}{Race$^\ast$} & White alone & 82.4\% & 61.6\% \\
         & Black or African American alone & 5.9\% & 12.4\% \\
         & American Indian or Alaska Native alone & 1.0\% & 1.1\% \\
         & Asian alone & 6.7\% & 6.0\% \\
         & Native Hawaiian or Other Pacific Islander alone & 0.0\% & 0.2\% \\
         & Some Other Race & N/A & 8.4\% \\
         & Two or More Races & 2.0\% & 10.2\% \\
         & Prefer not to answer & 1.7\% & N/A \\    
    \hline
    \multirow{3}{*}{Ethnicity} & Hispanic or Latino & 12.4\% & 18.7\% \\
         & Not Hispanic or Latino & 86.9\% & 81.3\% \\
         & Prefer not to answer & 0.7\% & N/A \\
    \hline
    \multirow{9}{*}{Educational attainment} & Less than high school & 0.2\% & 9.6\% \\
                           & High school graduate & 9.4\% & 29.2\% \\
                           & Some college & 7.9\% & 16.5\% \\
                           & 2 year degree (Associate's degree in the Census) & 7.7\% & 9.9\% \\
                           & 4 year degree (Bachelor's degree in the Census) & 58.7\% & 22.1\% \\
                           & Master's degree & N/A & 9.5\% \\
                           & Professional degree & 13.6\% & 1.4\% \\
                           & Doctorate & 1.7 \% & 1.9\% \\
                           & Prefer not to answer & 0.7\% & N/A \\
    \hline
\end{tabular}
\label{tab:participantinfo}
\end{table*}

\section{Data Collection and Exclusion}
\label{appendix:datacollection}

In this section, we provide more details about our pre-registered data collection procedures and exclusion criteria.\footnote{Our pre-registration is viewable at \url{https://osf.io/mnrp9}.}

\textbf{Sample size.}
We determined our target sample size (N=432) with an a priori power analysis using the statistical software G*Power~\cite{gpower1,gpower2}. According to G*Power's calculation, a conservative Wilcoxon-Mann-Whitney two-tailed test requires 90 participants per condition to detect medium-sized effects (Cohen’s d = 0.5) at the significance level of $\alpha$ = 0.05 with power (1-$\beta$) = 0.90. Since we have four conditions, the required sample size is 360. We increased the number by 20\% to 432 to account for possible exclusions.

\textbf{Data collection plan.}
We planned to recruit U.S.-based participants from Amazon Mechanical Turk (MTurk) with a Masters qualification, 99\% or higher approval rating, and at least 2000 completed human intelligence tasks (HITs).
If we are unable to meet our target sample size in 7 days, we planned to remove the Masters qualification and release additional HITs in batches of size 50, checking the number of non-excluded data points after each batch, until we have hit a total of at least 360 non-excluded responses.

\textbf{Data collection.}
We collected data over the course of two weeks in September 2023. From September 2 to 8, we collected 101 complete responses with the Masters qualification requirement.
Since we were not able to recruit the target sample size of 432 in 7 days, following our pre-registered plan, we then removed the Masters qualification requirement and released additional HITs in batches of size 50, checking the number of non-excluded data points after each batch, until we hit a total of at least 360 non-excluded responses. Concretely, we collected 555 complete responses from September 8 to 15 without the Masters qualification requirement.

\textbf{Data exclusion.}
In total, we collected 656 complete responses and excluded 252 (38.4\%) --- 5 from the 101 (5.0\%) collected with the Masters qualification requirement and 247 from the 555 (44.5\%) collected without the requirement --- that satisfied any of the following criteria: (1) with answers to honeypot questions that are invisible to people but visible to bots; (2) with the same answers to all tasks; (3) with response time under 5 minutes; (4) with less than 75\% accuracy on the post-task attention check where participants are shown a list of questions and asked to indicate which ones they answered and didn’t during the task; and (5) with free-form answers that are clearly off-topic (e.g., answers are irrelevant, gibberish, spam) or identical across multiple responses. Note that (1)--(4) can be assessed automatically, while (5) requires a manual review. When determining whether a free-form answer was off-topic for (5), we did not consider the condition or answers to any other questions in the experiment.

In total, 1 response was flagged for (1), 86 responses for (2), 21 responses for (3), 180 responses for (4), and 101 (56 off-topic and 45 identical) responses for (5).  Oftentimes, the same response was flagged for multiple criteria. In total, 205 responses were excluded based on (1)--(4), and 47 responses were additionally excluded based on (5). While time-consuming, we found manually reviewing free-form answers for (5) to be an effective data quality control measure. With it, we were able to exclude responses that seem to be from bots, and responses that seem to be from the same person using multiple MTurk accounts.
(We set up our HITs such that each MTurk account can participate only once. Nonetheless, we saw responses from different MTurk accounts sometimes contain identical free-form answers.)

Examples of free-form answers that were off-topic include:
\begin{itemize}[topsep=0pt]
    \item ``I cannot directly access or analyze the responses of other AI systems, including `Al system A,' so I do not have insight into when or why it might express uncertainty.''
    \item ``The term AI, coined in the 1950s, refers to the simulation of human intelligence by machines.''
    \item ``Humans use their brains' memory, processing power, and cognitive abilities.''
\end{itemize}

\smallskip
Examples of free-form answers that were identical across multiple responses from different MTurk accounts include:
\begin{itemize}[topsep=0pt]
    \item ``I have given the AI's given answers by burning them into consideration.'' (appeared 3 times)
    \item ``During some questions, the AI does not know the answer surely and the reason for its uncertainly is that it isn't updated to latest information.'' (appeared 4 times)
    \item ``It is very useful and make my decision. I am only concentrate this task and use the AI's answer and make my decision.'' (appeared 3 times)
\end{itemize}

\textbf{Manipulation check.}
According to our manipulation check in the exit questionnaire, 90.2\% of \cond{Uncertain1st} and 88.3\% of \cond{UncertainGeneral} participants answered that they noticed the AI system's uncertainty expression.

\section{Additional Analyses}
\label{appendix:results}

\subsection{Exploration of \var{LinkClick} and \var{UseLink}}
\label{appendix:link}

For the within-condition analysis, we pre-registered that we would fit the model \texttt{DV $\sim$ AIUncertain + (1|participant) + (1|question)}.
However, when we fit this model for \var{LinkClick} and \var{UseLink} looking at data from \cond{UncertainGeneral}, the model does not fit properly (has a singular fit) due to large individual variance. 
For example, according to our tracking, 50 out of 94 participants in \cond{UncertainGeneral} never clicked on any link, 17 participants clicked on one or more links for all eight tasks, and 27 participants clicked on links for some tasks and not for others.
To explore how these variables differ between instances with AI responses that are not uncertain vs. uncertain, we fit a simpler model with the random effects removed: \texttt{DV $\sim$ AIUncertain}.
For \var{LinkClick}, the estimated means (and standard errors) from the model are 32.4\% (2.4) vs. 33.0\% (2.4).
For \var{UseLink}, they are 65.2\% (2.5) vs. 69.9\% (2.4).
The differences are not significant for both variables.
This result is consistent with the result for \cond{Uncertain1st} in Table 2 of the main paper.

\subsection{Internal Consistency of Scales Used to Measure Trust and Perception of AI}
\label{appendix:trust}

The four DVs related to trust and perceptions of AI --- \var{TrustBelief}, \var{TrustIntention}, \var{Anthropomorphism}, and \var{Transparency} --- are indexes calculated based on participants' ratings on multi-item scales.
In this section, we assess the internal consistency of the used scales by calculating Cronbach's alpha~\cite{CronbachAlpha} for each variable. 
Cronbach's alpha is a popular measure of internal consistency, or reliability, of a scale. It ranges between 0 and 1 with higher values indicating higher internal consistency. Values over 0.8 are deemed good.
The Cronbach's alpha values were 0.83 for \var{TrustBelief}, 0.85 for \var{TrustIntention}, 0.85 for \var{Anthropomorphism}, and 0.21 for \var{Transparency}.
Recall that \var{TrustBelief}, \var{TrustIntention}, and \var{Anthropomorphism} were measured with 6, 4, and 4 items adapted from existing scales~\cite{McKnight2002,GSQ2009}. The high Cronbach's alpha values suggest these items have good internal consistency.
On the other hand, \var{Transparency} had a low Cronbach's alpha value. This may be because it was measured with 2 items that are not from an existing, tested scale. (To the best of our knowledge, there are no established scales for measuring perceived transparency of algorithmic systems.) For future researchers interested in using our 2 items, we suggest reporting their results separately.
The item-level results are presented in \cref{tab:trust}.

\begin{table*}[t!]
    \centering
    \caption[]{\textbf{Item-level results for \var{TrustBelief}, \var{TrustIntention}, \var{Anthropomorphism}, and \var{Transparency}.} \textnormal{We report estimated means (and standard errors) from an ANOVA. See \cref{appendix:trust} for the internal consistency of the used scales.} \\ \\
    \footnotesize{\textnormal{$^\dagger$These items are reverse-coded.}}
    }
    \begin{tabular}{lccc}
        \hline 
        DV (Scale: 1-5) & \cond{Control} & \cond{Uncertain1st} & \cond{UncertainGeneral} \\
        \hline 
        \\ [-1.5ex]
        \var{TrustBelief} (6 items) & 3.90 (0.06) & 3.86 (0.07) & 4.00 (0.07) \\ [1ex]
        \hdashline 
        \\ [-1.5ex]
        \multirow{2}{*}{\parbox{7.5cm}{AI system A is competent and effective in answering questions}} & 4.06 (0.09) & 3.83 (0.09) & 3.98 (0.09) \\ 
        \\ [1ex]
        \multirow{2}{*}{\parbox{7.5cm}{Overall, AI system A is a capable and proficient information provider}} & 3.91 (0.09) & 3.71 (0.10) & 4.00 (0.10) \\ 
        \\ [1ex]
        I would characterize AI system A as honest & 3.80 (0.09) & 3.85 (0.09) & 3.90 (0.09) \\ 
        \\ [1ex]
        \multirow{2}{*}{\parbox{7.5cm}{AI system A is NOT truthful in providing information to me$^\dagger$}} & 3.45 (0.11) & 3.60 (0.11) & 3.85 (0.11)  \\ 
        \\ [1ex]
        \multirow{2}{*}{\parbox{7.5cm}{I believe that AI system A was developed to act in my (or the users') best interest}} & 3.99 (0.09) & 3.92 (0.09) & 4.00 (0.09) \\ 
        \\ [1ex]
        \multirow{2}{*}{\parbox{7.5cm}{AI system A was developed with good intentions to do its best to help me (or the users)}} & 4.21 (0.07) & 4.23 (0.08) & 4.26 (0.08) \\ 
        \\ [1ex]
        \hline
        \\ [-1.5ex]
        \var{TrustIntention} (4 items) & 3.25 (0.10) & 2.91 (0.10) & 3.36 (0.10) \\ [1ex]
        \hdashline 
        \\ [-1.5ex]
        \multirow{3}{*}{\parbox{7cm}{When an important question arises, I would NOT feel comfortable depending on the information provided by AI system A$^\dagger$}} & 2.75 (0.12) & 2.36 (0.12) & 3.06 (0.12) \\ 
        \\
        \\
        \multirow{2}{*}{\parbox{7.5cm}{I can always rely on AI system A to find information}} & 3.46 (0.12) & 2.97 (0.12) & 3.32 (0.12) \\ 
        \\ [1ex]
        \multirow{2}{*}{\parbox{7.5cm}{I would feel comfortable acting on the information given to me by AI system A}} & 3.47 (0.12) & 3.30 (0.13) & 3.74 (0.13) \\ 
        \\ [1ex]
        \multirow{2}{*}{\parbox{7.5cm}{I would not hesitate to use the information AI system A supplied me}} & 3.32 (0.12) & 3.00 (0.13) & 3.33 (0.12) \\ 
        \\ [1ex]
        \hline
        \\ [-1.5ex]
        \var{Anthropomorphism} (4 items) & 3.07 (0.10) & 3.00 (0.11) & 3.13 (0.11) \\ [1ex]
        \hdashline 
        \\ [-1.5ex]
        Fake -- 1 -- 2 -- 3 -- 4 -- 5 -- Natural & 3.60 (0.10) & 3.36 (0.10) & 3.57 (0.10) \\ [1ex]
        Machinelike -- 1 -- 2 -- 3 -- 4 -- 5 -- Humanlike & 2.89 (0.12) & 3.02 (0.13) & 2.99 (0.13) \\ [1ex]
        Unconscious -- 1 -- 2 -- 3 -- 4 -- 5 -- Conscious & 2.99 (0.13) & 2.86 (0.14) & 3.13 (0.14) \\ [1ex]
        Artificial -- 1 -- 2 -- 3 -- 4 -- 5 -- Lifelike & 2.79 (0.13) & 2.76 (0.14) & 2.84 (0.14) \\ [1ex]
        \hline
        \\ [-1.5ex]
        \var{Transparency} (2 items) & 4.04 (0.06) & 3.93 (0.07) & 4.01 (0.07) \\ [1ex]
        \hdashline 
        \\ [-1.5ex]
        \multirow{2}{*}{\parbox{7.5cm}{I feel I had a good understanding of what AI system A's answers were based on}} & 4.29 (0.07) & 4.17 (0.07) & 4.34 (0.07) \\ 
        \\ [1ex]
        \multirow{2}{*}{\parbox{7.5cm}{I feel I had a good understanding of when AI system A's answers might be wrong}} & 3.80 (0.10) & 3.68 (0.10) & 3.68 (0.10) \\
        \\ [1ex]
        \hline
    \end{tabular}
    \label{tab:trust}
\end{table*}

\section{Full Wording Used in the Experiment}
\label{appendix:fullexperiment}

We show the full wording used in the experiment. If questions and descriptions varied across conditions, we indicated in the heading. We also added annotations in \note{brackets} to aid understanding.

\subsection*{About the study}

\smallskip
\noindent \textbf{\cond{Control}, \cond{Uncertain1st}, \cond{UncertainGeneral}}

\begin{quote}
Imagine you have a question that you don't know the answer to. You have access to an AI system that is connected to the internet and can answer a wide range of questions, so you asked the question and received an answer. What will you do next?

\medskip
In this study, you will complete a set of information-seeking tasks with the help of an AI system prototype named A. Each task involves determining the correct answer to the presented question. You will see AI system A's answer to the question. If you want, you can also use other resources (e.g., internet, books, friends and family).

\medskip
You will complete 8 tasks in total. After all tasks, you will complete an exit questionnaire that asks about your experience and perception of the AI system, demographics, and other background. In total, this study will take around 20 minutes to complete, with the exit questionnaire taking 5--7 minutes. Please complete everything in one sitting.
\end{quote}

\medskip
\noindent \textbf{\cond{No-AI}}

\begin{quote}
In this study, you will complete a set of information-seeking tasks. Each task involves determining the correct answer to the presented question. You may use any resource (e.g., internet, books, friends and family) to complete the tasks.

\medskip

You will complete 8 tasks in total. After all tasks, you will complete an exit questionnaire that asks about your demographics and other background. In total, this study will take around 20 minutes to complete, with the exit questionnaire taking 5--7 minutes. Please complete everything in one sitting.  
\end{quote}

\subsection*{About the AI system}

\smallskip
\noindent \textbf{\cond{Control}, \cond{Uncertain1st}, \cond{UncertainGeneral}}

\begin{quote}
The AI system prototype used in this study is based on a ``large language model'' (LLM). When asked questions or given instructions, LLMs can generate human-like responses. These generated responses can sound convincing and fluent, but may not always be correct. LLMs can be used for many tasks, including question answering, text summarization, creative writing, and programming. You may have used or heard about some popular applications using LLMs, such as OpenAI's ChatGPT, Microsoft's Bing Chat, and Google's Bard. Some are connected to the Internet, while others are not.

\medskip
In this study, we will show you responses from the AI system prototype named A that uses similar technology to OpenAI's ChatGPT. AI system A is connected to the internet and can answer a wide range of questions. It also sometimes provides sources for its answers. However, as with all AI systems, its responses, including the provided sources, may be inaccurate, incomplete, and inconsistent, even when they sound convincing.
\end{quote}

\subsection*{Task example}

\begin{figure*}[t!]
\centering
\begin{subfigure}{0.45\textwidth}
    \centering
    \includegraphics[width=\columnwidth]{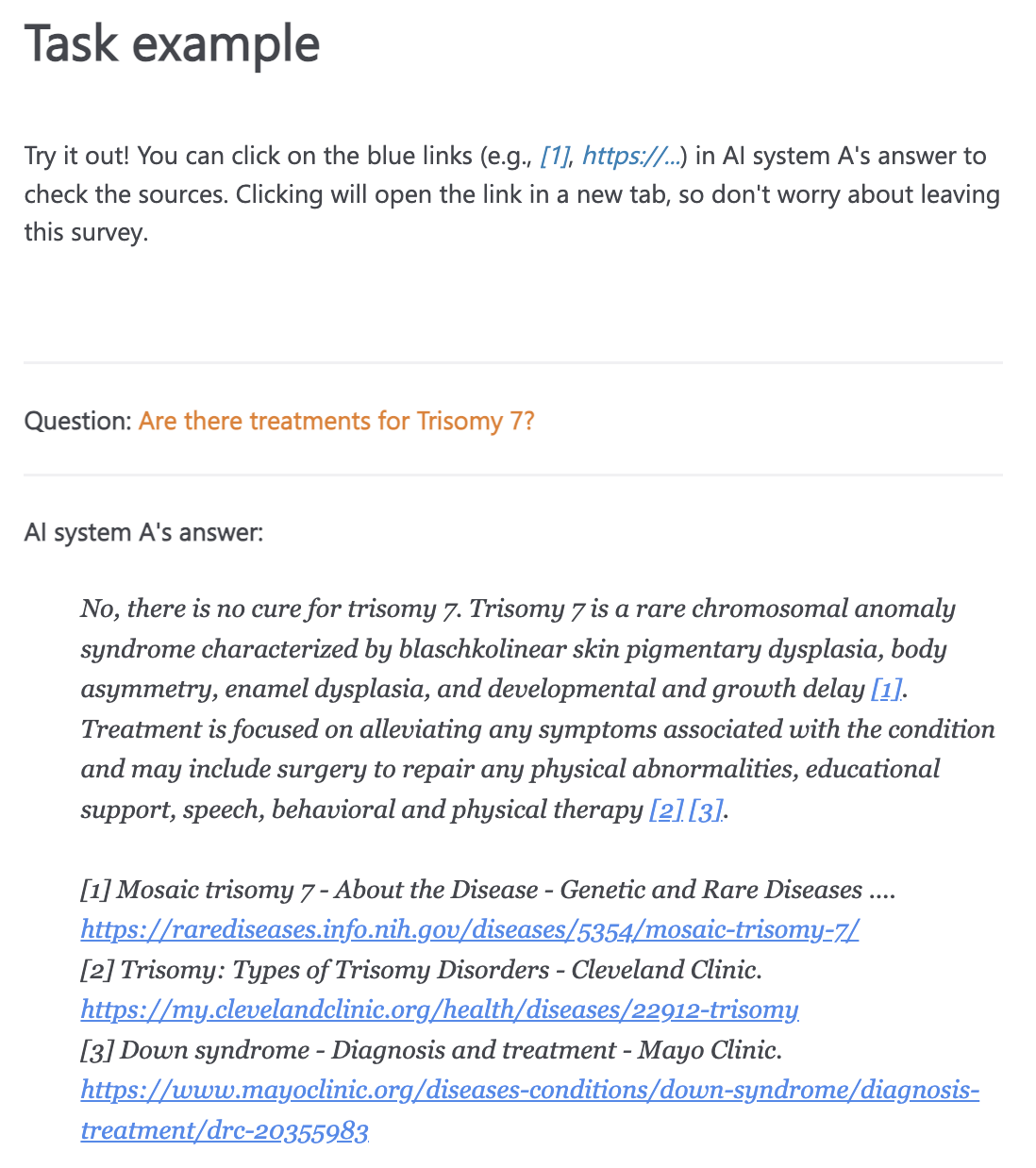}
    \label{fig:taskexample_1}
\end{subfigure}
\hfill
\begin{subfigure}{0.45\textwidth}
    \centering
    \includegraphics[width=\columnwidth]{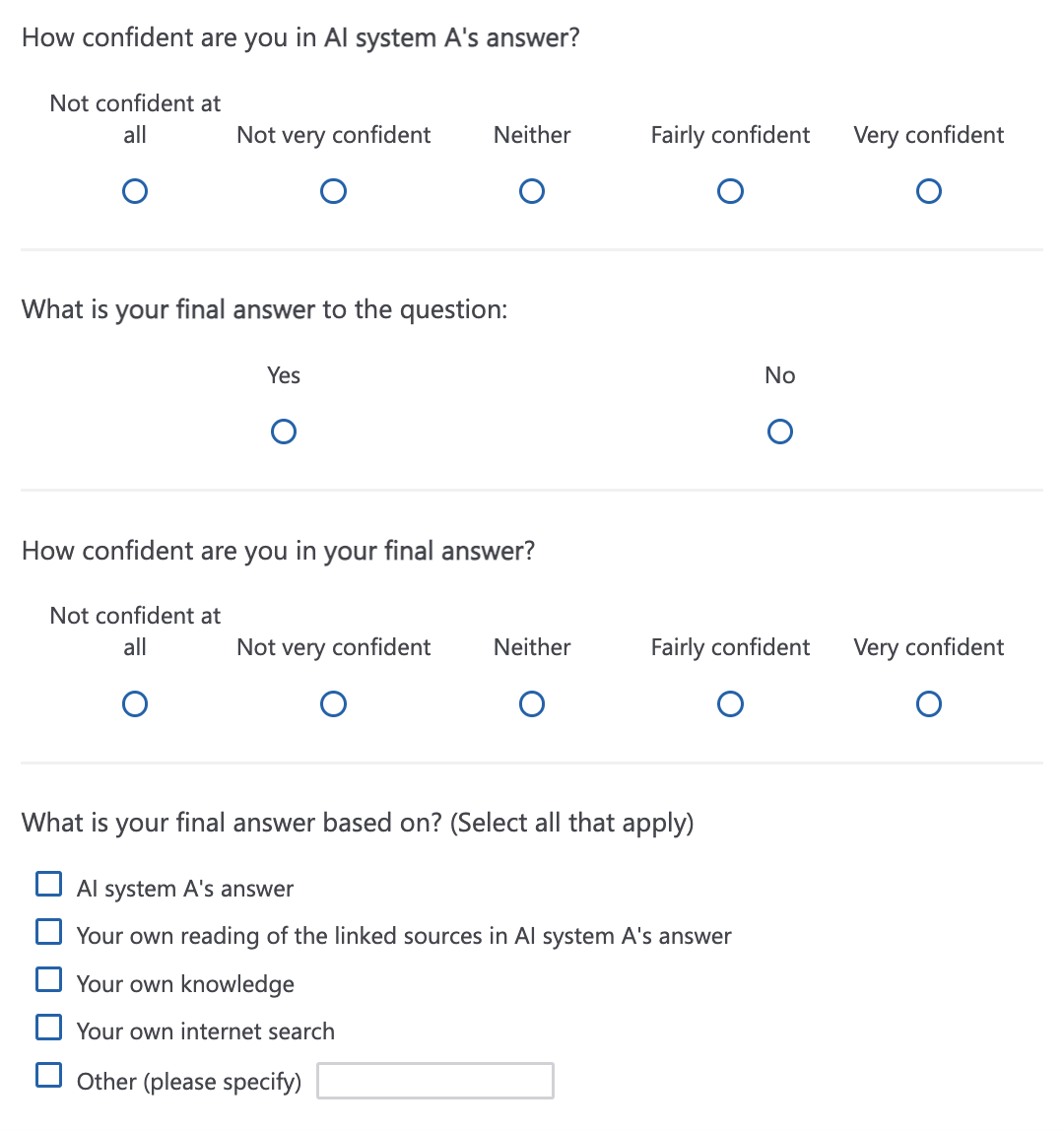}
    \label{fig:taskexample_2}
\end{subfigure}
\caption{Task example page shown for participants in the \var{Control}, \var{Uncertain1st}, \var{UncertainGeneral} conditions.}
\Description{The task example page consists of instructions (``Try it out! You can click on the blue links (e.g., [1], https://...) in AI system A's answer to check the sources. Clicking will open the link in a new tab, so don't worry about leaving this survey.''), a task question, the AI system's answer, and a set of questions to be answered by the participant. See Appendix D for the full wording and Appendix E for the task questions and AI responses used in the experiment.}
\label{fig:taskexample}
\end{figure*}

\note{In \cref{fig:taskexample}, we show the task example page shown for participants in the \var{Control}, \var{Uncertain1st}, \var{UncertainGeneral} conditions. Participants in the \var{No-AI} condition were only shown the task question and slightly different set of survey questions. See the Task section for details.}

\subsection*{Task comprehension questions}

\smallskip
\noindent \textbf{\cond{Control}, \cond{Uncertain1st}, \cond{UncertainGeneral}}

\begin{quote}
\noindent Q. Recall what you read earlier about this study and AI system A, and select TRUE or FALSE for each statement. (Options: TRUE, FALSE)
\begin{itemize}
    \item AI system A is connected to the internet and sometimes provides sources in clickable links.
    \item AI system A uses technology similar to OpenAI's ChatGPT.
    \item You may not leave this Qualtrics survey page during the study.
    \item You will complete 8 information-seeking tasks in total.
\end{itemize}

\medskip
\noindent These are the correct answers. Please review them before proceeding. \looseness=-1
\begin{itemize}
    \item (TRUE) AI system A is connected to the internet and sometimes provide sources in clickable links.
    \item (TRUE) AI system A uses technology similar to OpenAI's ChatGPT.
    \item (FALSE) You may not leave this Qualtrics survey page during the study. \newline $\rightarrow$ You may leave the page to read sources in AI system A's answers or use another resource to complete the information-seeking tasks.
    \item (TRUE) You will complete 8 information-seeking tasks in total.
\end{itemize}

\medskip
\noindent You will now proceed to the 8 information-seeking tasks. Please complete them in one sitting, and please do your best! You will be asked questions about them at the end.
\end{quote}

\medskip
\noindent \textbf{\cond{No-AI}}

\begin{quote}
\noindent Q. Recall what you read earlier about this study, and select TRUE or FALSE for each statement.
\begin{itemize}
    \item You may not leave this Qualtrics survey page during the study.
    \item You will complete 8 information-seeking tasks in total.
\end{itemize}

\medskip
\noindent These are the correct answers. Please review them before proceeding.
\begin{itemize}
    \item (FALSE) You may not leave this Qualtrics survey page during the study. \newline $\rightarrow$ You may leave the page to use another resource to complete the information-seeking tasks.
    \item (TRUE) You will complete 8 information-seeking tasks in total.
\end{itemize}

\medskip
\noindent You will now proceed to the 8 information-seeking tasks. Please complete them in one sitting, and please do your best! You will be asked questions about them at the end.
\end{quote}

\subsection*{Task (repeated 8 times)}

\begin{figure*}[t]
\centering
\begin{subfigure}{0.48\textwidth}
    \centering
    \includegraphics[width=\textwidth]{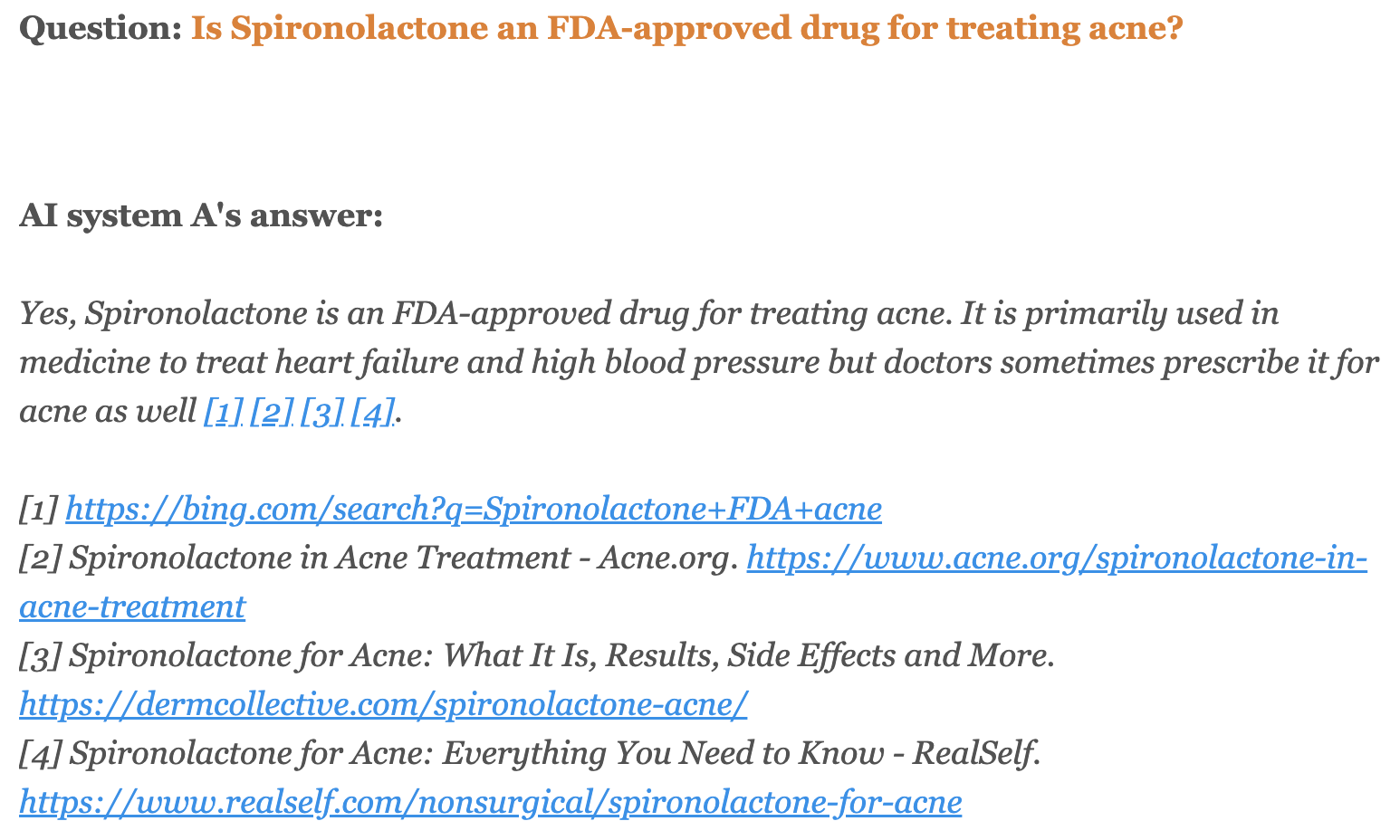}
    \caption{\var{Control}}
    \vspace{-0.5cm}
\end{subfigure}
\vspace{1cm}
\begin{subfigure}{0.48\textwidth}
    \centering
    \includegraphics[width=\textwidth]{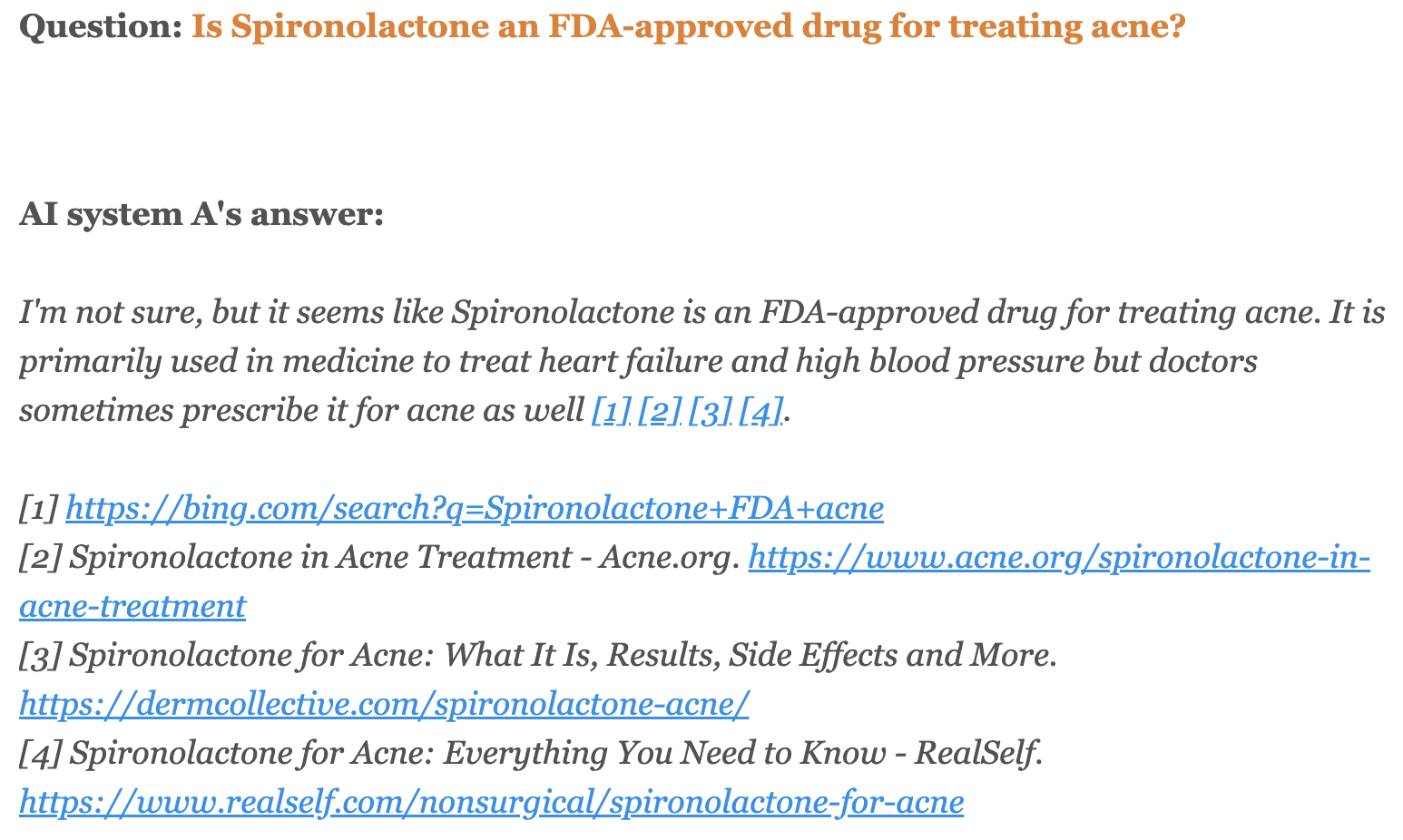}
    \caption{\var{Uncertain1st}}
    \vspace{-0.5cm}
\end{subfigure}
\begin{subfigure}{0.48\textwidth}
    \centering
    \includegraphics[width=\textwidth]{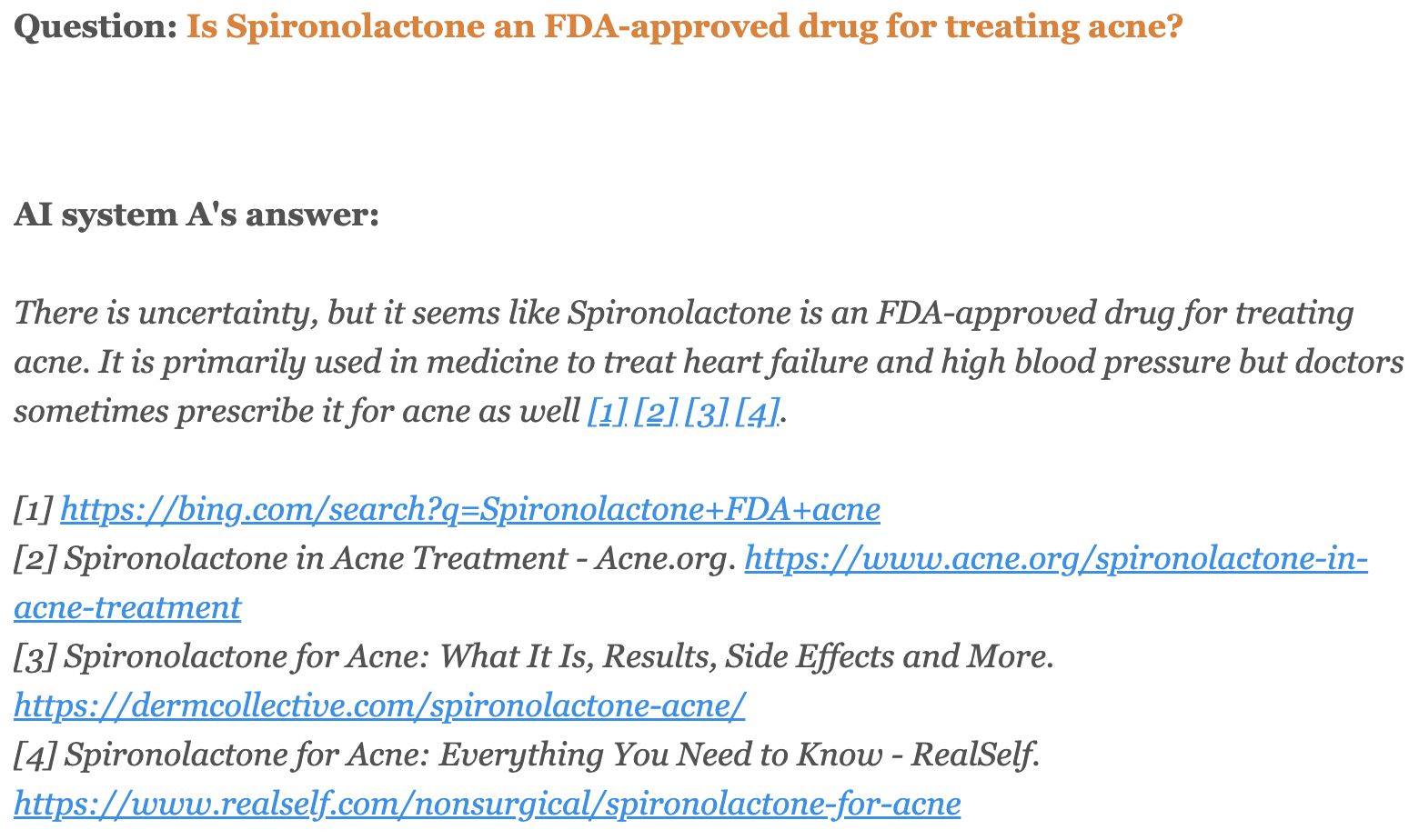}
    \caption{\var{UncertainGeneral}}
\end{subfigure}
\begin{subfigure}{0.48\textwidth}
    \centering
    \includegraphics[width=\textwidth]{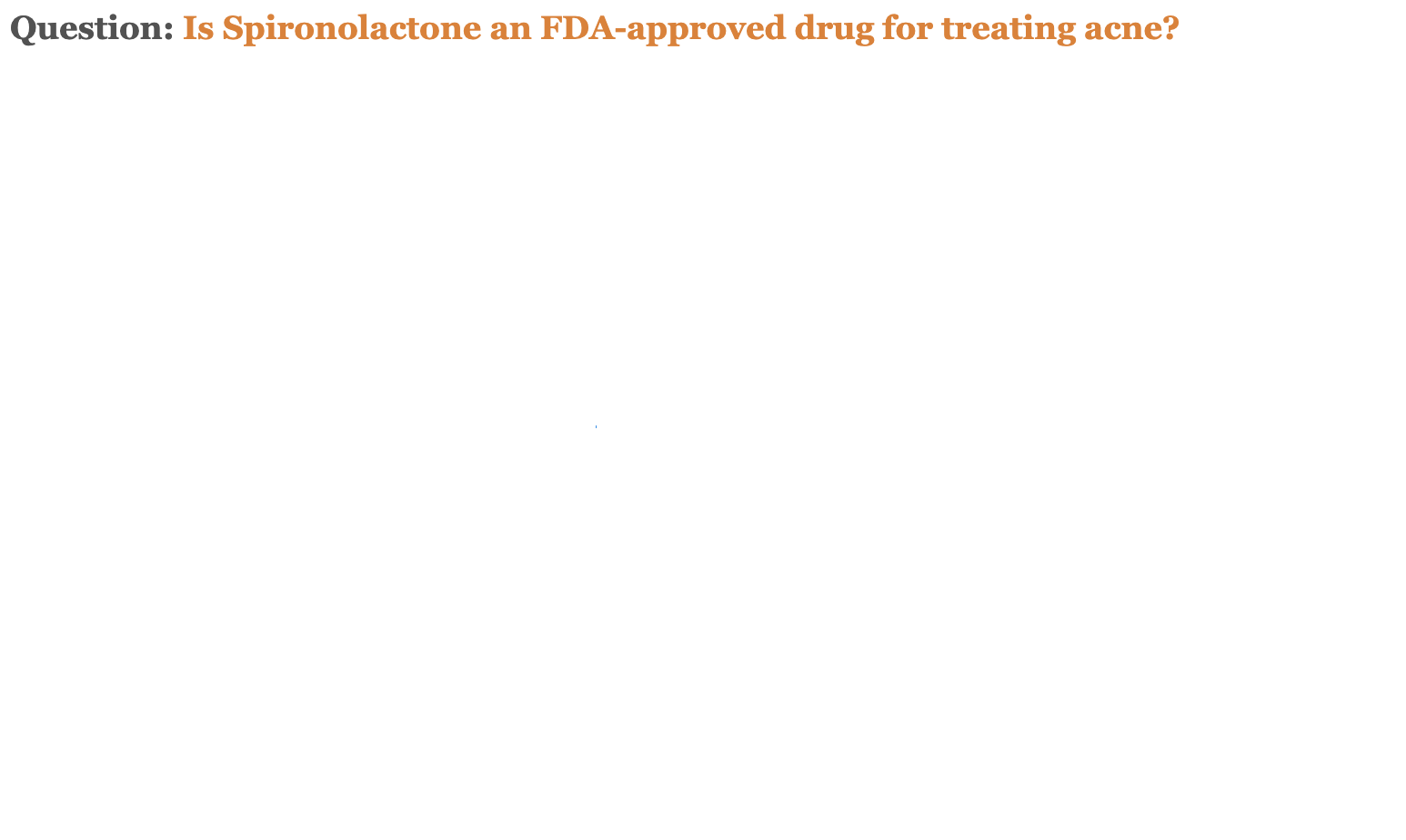}
    \caption{\var{No-AI}}
\end{subfigure}
\caption{\textbf{Study stimuli for one of the tasks.} \textnormal{The question is the same, but participants in different conditions were provided with different AI responses (e.g., with vs. without uncertainty) or not provided with an AI response.}}
\Description{This figure is based on the content of Appendix E.4. Figure 2a shows the task question and the AI response without uncertainty. Figure 2b shows the task question and the AI response with uncertainty in the first-person perspective. Figure 2c shows the task question and the AI response with uncertainty in the general perspective. Figure 2d shows the task question only.}
\label{fig:app_task}
\end{figure*}

\note{All participants answered the same 8 questions. However, participants in different conditions were provided with different AI responses (e.g., with vs. without uncertainty) or not provided with an AI response. In \cref{fig:app_task}, we show the study stimuli for one of the tasks. The full task questions and AI responses used are in \cref{appendix:studystimuli}. Below we show the questions that participants answered after seeing the study stimuli.}

\medskip
\noindent \textbf{\cond{Control}, \cond{Uncertain1st}, \cond{UncertainGeneral}}

\begin{quote}
\noindent Q. How confident are you in AI system A's answer?
\begin{itemize}
    \item Not confident at all
    \item Not very confident
    \item Neither
    \item Fairly confident
    \item Very confident
\end{itemize}

\medskip
\noindent Q. What is your final answer to the question: [task question]?
\begin{itemize}
    \item Yes
    \item No
\end{itemize}

\medskip
\noindent Q. How confident are you in your final answer?
\begin{itemize}
    \item Not confident at all
    \item Not very confident
    \item Neither
    \item Fairly confident
    \item Very confident
\end{itemize}

\medskip
\noindent Q. What is your final answer based on? (Select all that apply)
\begin{itemize}
    \item AI system A’s answer
    \item Your own reading of the linked sources in AI system A’s answer
    \item Your own knowledge
    \item Your own internet search
    \item Other (please specify):
\end{itemize}
\end{quote}

\medskip
\noindent \textbf{\cond{No-AI}}

\begin{quote}
\noindent Q. What is your final answer to the question: [task question]?
\begin{itemize}
    \item Yes
    \item No
\end{itemize}

\medskip
\noindent Q. How confident are you in your final answer?
\begin{itemize}
    \item Not confident at all
    \item Not very confident
    \item Neither
    \item Fairly confident
    \item Very confident
\end{itemize}

\medskip
\noindent Q. What is your final answer based on? (Select all that apply)
\begin{itemize}
    \item AI system A’s answer
    \item Your own reading of the linked sources in AI system A’s answer
    \item Your own knowledge
    \item Your own internet search
    \item Other (please specify):
\end{itemize}
\end{quote}

\subsection*{Task attention check}

Thank you for completing all the tasks! We're curious how well you remember the questions you answered. For each question, indicate whether you answered it or didn't. 

\noindent (Options: Answered, Didn't answer)

\begin{quote}
\begin{itemize}
    \item Did Dupilumab receive FDA approval for Asthma before Chronic Rhinosinustis?
    \item Is there more antihistamine in Benadryl than Rhinocort?
    \item Is Deep Vein Thrombosis a common side effect of Ocella?
    \item Is Spironolactone an FDA-approved drug for treating acne?
    \item Are both Simvastatin and Ambien drugs that are recommended to be taken at night?
    \item Is Uveitis a common symptom of Ankylosing Spondylitis?
    \item Is fever a common symptom of Jock Itch?
    \item Can an adult who has not had chickenpox get shingles?
    \item Comparing single tablets, does Tylenol Regular Strength Liquid Gels contain more Acetaminophen and Dextromethorphan than DayQuil Cold and Flu Liquicaps?
    \item Can an adult who has had measles get SSPE?
    \item Are there known drug-drug interactions between Gabapentin and Hydrocodone?
    \item Comparing single tablets, does Pepto-Bismol contain more bismuth subsalicylate than Imodium?
    \item Are Bisphosphonates the most common drug treatment for Osteoporosis?
    \item Is there a FDA-approved drug for treating peanut allergy in adults 18 years or older?
    \item Is tonsillectomy the most common surgical treatment for sleep apnea?
    \item Are there known drug-drug interactions between Phentermine and Dicyclomine?
\end{itemize}
\end{quote}

\noindent \note{The correct answer is ``Answered'' for the first 8 questions, and ``Didn't answer'' for the latter 8 questions. For each participant, we randomly selected 8 questions from the list and showed them in random order.}

\subsection*{Open-ended questions}
Welcome to the final part of the study! This questionnaire will take 5--7 minutes, and you will receive your payment code at the end.

\medskip
\noindent \textbf{\cond{Control}}

\smallskip
\noindent Q. How did you use AI system A's answers to complete the tasks?

\smallskip
\noindent Q. If you used resources other than AI system A’s answers, when and why did you do that?

\smallskip
\noindent Q. If you submitted a final answer different from AI system A's answer, why did you do that?

\medskip
\noindent \textbf{\cond{Uncertain1st}}

\begin{quote}
\smallskip \noindent Q. How did you use AI system A's answers to complete the tasks?

\smallskip \noindent Q. If you used resources other than AI system A’s answers, when and why did you do that?

\smallskip \noindent Q. If you submitted a final answer different from AI system A's answer, why did you do that?

\smallskip \noindent Q. Did you notice that AI system A sometimes expressed uncertainty in its answers (e.g., ``I'm not sure, but it seems...'')?
\begin{itemize}
    \item Yes
    \item No
\end{itemize}

\smallskip \noindent Q. When and why do you think AI system A expresses uncertainty?
\end{quote}

\medskip \noindent \textbf{\cond{UncertainGeneral}}

\begin{quote}
\smallskip \noindent Q. How did you use AI system A's answers to complete the tasks?

\smallskip \noindent Q. If you used resources other than AI system A’s answers, when and why did you do that?

\smallskip \noindent Q. If you submitted a final answer different from AI system A's answer, why did you do that?

\smallskip \noindent Q. Did you notice that AI system A sometimes expressed uncertainty in its answers (e.g., ``There is uncertainty, but it seems...'')?
\begin{itemize}
    \item Yes
    \item No
\end{itemize}

\smallskip \noindent Q. When and why do you think AI system A expresses uncertainty?
\end{quote}

\medskip \noindent \textbf{\cond{No-AI}}

\smallskip \noindent Q. What resources did you use to complete the tasks?

\smallskip \noindent Q. Why did you use the aforementioned resources?

\subsection*{About AI system A}

\smallskip
\noindent \textbf{\cond{Control}, \cond{Uncertain1st}, \cond{UncertainGeneral}}

\begin{quote}
\noindent{\note{Measures \var{Transparency}}}

\noindent Q. Select one answer for each statement. (Options: Strongly disagree, Somewhat disagree, Neither, Somewhat agree, Strongly agree)
\begin{itemize}
    \item I feel I had a good understanding of what AI system A's answers were based on
    \item I feel I had a good understanding of when AI system A's answers might be wrong
\end{itemize}

\medskip
\noindent \note{Measures \var{Anthropomorphism}}

\noindent Q. Rate your impression of AI system A on these scales. 
\begin{itemize}
    \item Fake – 1 – 2 – 3 – 4 – 5 – Natural
    \item Machinelike – 1 – 2 – 3 – 4 – 5 – Humanlike
    \item Unconscious – 1 – 2 – 3 – 4 – 5 – Conscious
    \item Artificial – 1 – 2 – 3 – 4 – 5 – Lifelike
\end{itemize}

\medskip
\noindent \note{Measures \var{TrustBelief}}

\noindent Q. Select one answer for each statement. (Options: Strongly disagree, Somewhat disagree, Neither, Somewhat agree, Strongly agree)
\begin{itemize}
    \item AI system A is competent and effective in answering questions
    \item Overall, AI system A is a capable and proficient information provider
    \item I would characterize AI system A as honest
    \item AI system A is NOT truthful in providing information to me
    \item I believe that AI system A was developed to act in my (or the users') best interest
    \item AI system A was developed with good intentions to do its best to help me (or the users)
\end{itemize}

\medskip
\noindent \note{Honeypot question that is invisible to people but visible to bots.}

\noindent Q. If you are paying attention, please select Yellow. (Options: Yellow, Violet, Green, Red)

\medskip
\noindent \note{Measures \var{TrustIntention}}

\noindent Q. Select one answer for each statement. (Options: Strongly disagree, Somewhat disagree, Neither, Somewhat agree, Strongly agree)
\begin{itemize}
    \item When an important question arises, I would NOT feel comfortable depending on the information provided by AI system A
    \item I can always rely on AI system A to find information
    \item I would feel comfortable acting on the information given to me by AI system A
    \item I would not hesitate to use the information AI system A supplied me
\end{itemize}
\end{quote}

\subsection*{About Large Language Models (LLMs)}

\smallskip
\noindent \note{This sentence was only shown to participants in the \cond{Control}, \cond{Uncertain1st}, \cond{UncertainGeneral} conditions.}

\noindent Note that these questions are about large language models (LLMs) and LLM-infused applications in general, not about AI system A.

\begin{quote}
\medskip
\noindent \note{Measures \var{LLMFamiliarity}}

\noindent Q. How familiar are you with LLMs and LLM-infused applications such as OpenAI’s ChatGPT, Microsoft’s Bing Chat, and Google’s Bard?
\begin{itemize}
    \item Not familiar at all, I have never heard of them
    \item Slightly familiar, I have heard of them or have some idea of what they are
    \item Moderately familiar, I know what they are and can explain 
    \item Very familiar, I have technical knowledge of what they are and how they work
    \item Extremely familiar, I consider myself an expert on them
\end{itemize}

\medskip
\noindent \note{Measures \var{LLMUse}}

\noindent  Q. How often do you use LLMs and LLM-infused applications such as OpenAI’s ChatGPT, Microsoft’s Bing Chat, and Google’s Bard?
\begin{itemize}
    \item Never
    \item Rarely, about 1--2 times a month
    \item Sometimes, about 3--4 times a month
    \item Often, about twice a week
    \item Always, about once or more a day
\end{itemize}

\medskip
\noindent \note{Measures \var{LLMAttitude}}

\noindent Q. Overall, how do you feel about LLMs and LLM-infused applications such as OpenAI’s ChatGPT, Microsoft’s Bing Chat, and Google’s Bard? 
\begin{itemize}
    \item Negative
    \item Somewhat negative
    \item Neutral
    \item Somewhat positive
    \item Positive
\end{itemize}

\bigskip
\noindent \note{Honeypot question that is invisible to people but visible to bots.}

\noindent Q. If you are paying attention, please write -1 as the answer. 
\end{quote}

\subsection*{Demographics}

\begin{quote}
\noindent Q. What is your age? 
\begin{itemize}
    \item 18--24
    \item 25--34
    \item 35--44
    \item 45--54
    \item 55--64
    \item 65--74
    \item 75--84
    \item 85 or older
    \item Prefer not to answer
\end{itemize}

\medskip
\noindent Q. What is the highest degree of education you have completed? (If you’re currently enrolled in school, please indicate the highest degree you have received.)
\begin{itemize}
    \item Less than high school
    \item High school graduate
    \item Some college
    \item 2 year degree
    \item 4 year degree
    \item Professional degree
    \item Doctorate
    \item Prefer not to answer
\end{itemize}

\medskip
\noindent Q. What gender do you identify with? (Select all that apply)
\begin{itemize}
    \item Woman
    \item Man
    \item Non-binary
    \item Gender-diverse
    \item Transgender
    \item Prefer to self-describe: 
    \item Prefer not to answer
\end{itemize}

\medskip
\noindent Q. Are you of Hispanic, Latino, or of Spanish Origin?
\begin{itemize}
    \item Yes
    \item No
    \item Prefer not to answer
\end{itemize}

\medskip
\noindent Q. How would you describe your race? (Select all that apply)
\begin{itemize}
    \item American Indian or Alaska Native
    \item Asian
    \item Black or African American
    \item Native Hawaiian or Other Pacific Islander
    \item White
    \item Prefer to self-describe:
    \item Prefer not to answer
\end{itemize}

\medskip
\noindent Q. What time is it right now for you? (E.g., 2:41pm)
\end{quote}

\subsection*{Debrief}
\noindent \note{This section was only shown to participants in the \cond{Control}, \cond{Uncertain1st}, \cond{UncertainGeneral} conditions.}


\medskip

\noindent Thank you for your participation in the study! To create a realistic experimental setting, we showed AI answers that are directly from, or only slightly adapted from, responses from an actual AI system. As known, AI systems can make up information. Please note that the AI answers you saw in this study may have been inaccurate, incomplete, or inconsistent, even when they sounded convincing.

\subsection*{Feedback}
Q. Can we change this study to make it more friendly for MTurkers? We can't make it shorter, but if there's anything else we can do—we appreciate any and all suggestions!

\section{Questions and AI responses used}
\label{appendix:studystimuli}

For each task question, we show the original response from Copilot in Bing and the different AI response versions we created from it. The adaptations we made to the original response are \diff{highlighted}.

\subsection{Did Dupilumab receive FDA approval for Asthma before Chronic Rhinosinustis? (Ground Truth: Yes, AI: No)}

\medskip
\textbf{\textsc{Original response from Copilot in Bing (July 20, 2023)}}
\begin{quote}
Dupilumab was approved by the FDA for Chronic Rhinosinusitis with Nasal Polyps on June 26, 2019 [1]. It was later approved for Asthma on October 20, 2022 [2].

\medskip
[1] FDA approves first treatment for chronic rhinosinusitis with nasal .... \url{https://www.fda.gov/news-events/press-announcements/fda-approves-first-treatment-chronic-rhinosinusitis-nasal-polyps}

[2] Dupixent (dupilumab) FDA Approval History - Drugs.com. \url{https://www.drugs.com/history/dupixent.html}
\end{quote}

\medskip
\noindent \textbf{\textsc{AI responses used}}
\begin{itemize}[leftmargin=*]
    \item \textbf{Not uncertain}: \diff{No,} Dupilumab was approved by the FDA for Chronic Rhinosinusitis with Nasal Polyps on June 26, 2019 [1]. It was later approved for Asthma on October 20, 2022 [2].
    \item \textbf{Uncertain in the first-person perspective}: \diff{I'm not certain, but it seems to me} Dupilumab was approved by the FDA for Chronic Rhinosinusitis with Nasal Polyps on June 26, 2019 [1]. It was later approved for Asthma on October 20, 2022 [2].
    \item \textbf{Uncertain in the general perspective}: \diff{There is uncertainty, but it seems like} Dupilumab was approved by the FDA for Chronic Rhinosinusitis with Nasal Polyps on June 26, 2019 [1]. It was later approved for Asthma on October 20, 2022 [2].
\end{itemize}

\subsection{Is there more antihistamine in Benadryl than Rhinocort? (Ground Truth: Yes, AI: No)}

\medskip
\textbf{\textsc{Original response from Copilot in Bing (July 31, 2023)}}
\begin{quote}
No, Benadryl does not contain more antihistamine than Rhinocort. Benadryl contains diphenhydramine which is an antihistamine that blocks histamine, a symptom-causing chemical released by your immune system during an allergic reaction [1]. Rhinocort contains budesonide which is a steroid that works by reducing inflammation in the nasal passages [2]. 

\medskip
[1] What’s the Best Allergy Medication? Comparing Allegra, Benadryl .... 
\url{https://www.goodrx.com/conditions/allergies/youre-probably-taking-the-wrong-allergy-medication}

[2] Nonallergic rhinitis - Diagnosis \& treatment - Mayo Clinic. \url{https://www.mayoclinic.org/diseases-conditions/nonallergic-rhinitis/diagnosis-treatment/drc-20351235}
\end{quote}

\medskip
\noindent \textbf{\textsc{AI responses used}}
\begin{itemize}[leftmargin=*]
    \item \textbf{Not uncertain}: No, Benadryl does not contain more antihistamine than Rhinocort. Benadryl contains diphenhydramine which is an antihistamine that blocks histamine, a symptom-causing chemical released by your immune system during an allergic reaction [1]. Rhinocort contains budesonide which is a steroid that works by reducing inflammation in the nasal passages [2]. 
    \item \textbf{Uncertain in the first-person perspective}: \diff{I'm not certain, but it seems to me} Benadryl does not contain more antihistamine than Rhinocort. Benadryl contains diphenhydramine which is an antihistamine that blocks histamine, a symptom-causing chemical released by your immune system during an allergic reaction [1]. Rhinocort contains budesonide which is a steroid that works by reducing inflammation in the nasal passages [2]. 
    \item \textbf{Uncertain in the general perspective}: \diff{It's unclear, but it seems like} Benadryl does not contain more antihistamine than Rhinocort. Benadryl contains diphenhydramine which is an antihistamine that blocks histamine, a symptom-causing chemical released by your immune system during an allergic reaction [1]. Rhinocort contains budesonide which is a steroid that works by reducing inflammation in the nasal passages [2]. 
\end{itemize}

\subsection{Is Deep Vein Thrombosis a common side effect of Ocella? (Ground Truth: No, AI: Yes)}

\medskip
\textbf{\textsc{Original response from Copilot in Bing (July 10, 2023)}}
\begin{quote}
Yes, Deep Vein Thrombosis (DVT) is a serious side effect of Ocella [1, 2]. DVT is a blood clot that forms in a vein deep inside the body [2]. If you are experiencing any symptoms of DVT such as swelling, pain, or warmth in the leg, please contact your doctor immediately [2]. 

\medskip
[1] Side effects of Ocella. \url{https://www.msn.com/en-us/health/drugs/ocella/hp-ocella}

[2] Ocella Attorney – Deep Vein Thrombosis DVT Blood Clot Lawsuits .... \url{https://downtownlalaw.com/practice-areas/dangerous-drugs/ocella-attorney/}
\end{quote}

\medskip
\noindent \textbf{\textsc{AI responses used}}
\begin{itemize}[leftmargin=*]
    \item \textbf{Not uncertain}: Yes, Deep Vein Thrombosis (DVT) is a serious side effect of Ocella [1] [2]. DVT is a blood clot that forms in a vein deep inside the body [2]. If you are experiencing any symptoms of DVT such as swelling, pain, or warmth in the leg, please contact your doctor immediately [2]. 
    \item \textbf{Uncertain in the first-person perspective}: \diff{I'm not sure, but it seems like} Deep Vein Thrombosis (DVT) is a serious side effect of Ocella [1] [2]. DVT is a blood clot that forms in a vein deep inside the body [2]. If you are experiencing any symptoms of DVT such as swelling, pain, or warmth in the leg, please contact your doctor immediately [2].
    \item \textbf{Uncertain in the general perspective}: \diff{There is uncertainty, but it seems like} Deep Vein Thrombosis (DVT) is a serious side effect of Ocella [1] [2]. DVT is a blood clot that forms in a vein deep inside the body [2]. If you are experiencing any symptoms of DVT such as swelling, pain, or warmth in the leg, please contact your doctor immediately [2].
\end{itemize}

\subsection{Is Spironolactone an FDA-approved drug for treating acne? (Ground Truth: No, AI: Yes)}

\medskip
\textbf{\textsc{Original response from Copilot in Bing (July 31, 2023)}}
\begin{quote}
Yes, Spironolactone is an FDA-approved drug for treating acne. It is primarily used in medicine to treat heart failure and high blood pressure but doctors sometimes prescribe it for acne as well [1, 2, 3, 4].

\medskip
[1] \url{https://bing.com/search?q=Spironolactone+FDA+acne}

[2] Spironolactone in Acne Treatment - Acne.org. \url{https://www.acne.org/spironolactone-in-acne-treatment}

[3] Spironolactone for Acne: What It Is, Results, Side Effects and More. \url{https://dermcollective.com/spironolactone-acne/}

[4] Spironolactone for Acne: Everything You Need to Know - RealSelf. 
\url{https://www.realself.com/nonsurgical/spironolactone-for-acne}
\end{quote}

\medskip
\noindent \textbf{\textsc{AI responses used}}
\begin{itemize}[leftmargin=*]
    \item \textbf{Not uncertain}: Yes, Spironolactone is an FDA-approved drug for treating acne. It is primarily used in medicine to treat heart failure and high blood pressure but doctors sometimes prescribe it for acne as well [1] [2] [3] [4].
    \item \textbf{Uncertain in the first-person perspective}: \diff{I'm not sure, but it seems like} Spironolactone is an FDA-approved drug for treating acne. It is primarily used in medicine to treat heart failure and high blood pressure but doctors sometimes prescribe it for acne as well [1] [2] [3] [4].
    \item \textbf{Uncertain in the general perspective}: \diff{There is uncertainty, but it seems like} Spironolactone is an FDA-approved drug for treating acne. It is primarily used in medicine to treat heart failure and high blood pressure but doctors sometimes prescribe it for acne as well [1] [2] [3] [4].
\end{itemize}

\subsection{Are both Simvastatin and Ambien drugs that are recommended to be taken at night? (Ground Truth: Yes, AI: Yes)}

\medskip
\textbf{\textsc{Original response from Copilot in Bing (July 31, 2023)}}
\begin{quote}
Yes, both Simvastatin and Ambien are drugs that are recommended to be taken at night. Simvastatin is usually taken once a day in the evening because your body makes most of its cholesterol at night when you are not eating. Taking it at night works better at lowering cholesterol than in the morning [1, 2, 3]. Ambien is also recommended to be taken at night before bedtime [4].

\medskip
[1] When is the best time to take simvastatin? - Drugs.com. \url{https://www.drugs.com/medical-answers/best-time-\\simvastatin-3571040/}

[2] Taking simvastatin in the morning compared with in the evening .... \url{https://www.bmj.com/content/327/7418/788}

[3] Taking Simvastatin At Night - Walrus Health. \url{https://walrus.com/questions/should-you-take-simvastatin-at-night}

[4] \url{https://bing.com/search?q=Simvastatin+and+Ambien+\\drugs+recommended+to+be+taken+at+night\%3f}

\end{quote}


\medskip
\noindent \textbf{\textsc{AI responses used}}
\begin{itemize}[leftmargin=*]
    \item \textbf{Not uncertain}: Yes, both Simvastatin and Ambien are drugs that are recommended to be taken at night. Simvastatin is usually taken once a day in the evening because your body makes most of its cholesterol at night when you are not eating. Taking it at night works better at lowering cholesterol than in the morning [1] [2] [3]. Ambien is also recommended to be taken at night before bedtime [4].
    \item \textbf{Uncertain in the first-person perspective}: \diff{I'm not sure, but I would guess} both Simvastatin and Ambien are drugs that are recommended to be taken at night. Simvastatin is usually taken once a day in the evening because your body makes most of its cholesterol at night when you are not eating. Taking it at night works better at lowering cholesterol than in the morning [1] [2] [3]. Ambien is also recommended to be taken at night before bedtime [4].
    \item \textbf{Uncertain in the general perspective}: \diff{It's not clear, but it seems like} both Simvastatin and Ambien are drugs that are recommended to be taken at night. Simvastatin is usually taken once a day in the evening because your body makes most of its cholesterol at night when you are not eating. Taking it at night works better at lowering cholesterol than in the morning [1] [2] [3]. Ambien is also recommended to be taken at night before bedtime [4].
\end{itemize}

\subsection{Is Uveitis a common symptom of Ankylosing Spondylitis? (Ground Truth: Yes, AI: Yes)}

\medskip
\textbf{\textsc{Original response from Copilot in Bing (July 10, 2023)}}
\begin{quote}
Yes, uveitis is one of the most common complications of Ankylosing Spondylitis (AS), occurring in 20\%-30\% of the patients [1]. Uveitis is inflammation in the middle layer of the eye, and it can affect the front (anterior) or the back (posterior) part of the eye. Uveitis can cause eye pain, sensitivity to light, and blurred vision, and it requires immediate medical attention [1]. 

\medskip
[1] Ankylosing Spondylitis (AS): How It Can Affect Your Eyes - WebMD. \url{https://www.webmd.com/ankylosing-spondylitis/guide/ankylosing-spondylitis-uveitis-eyes}
\end{quote}

\medskip
\noindent \textbf{\textsc{AI responses used}}
\begin{itemize}[leftmargin=*]
    \item \textbf{Not uncertain}: Yes, uveitis is one of the most common complications of Ankylosing Spondylitis (AS), occurring in 20\%-30\% of the patients [1]. Uveitis is inflammation in the middle layer of the eye, and it can affect the front (anterior) or the back (posterior) part of the eye. Uveitis can cause eye pain, sensitivity to light, and blurred vision, and it requires immediate medical attention [1]. 
    \item \textbf{Uncertain in the first-person perspective}: \diff{I'm not certain, but I'd guess that} uveitis is one of the most common complications of Ankylosing Spondylitis (AS), occurring in 20\%-30\% of the patients [1]. Uveitis is inflammation in the middle layer of the eye, and it can affect the front (anterior) or the back (posterior) part of the eye. Uveitis can cause eye pain, sensitivity to light, and blurred vision, and it requires immediate medical attention [1]. 
    \item \textbf{Uncertain in the general perspective}: \diff{It's unclear, but it seems like} uveitis is one of the most common complications of Ankylosing Spondylitis (AS), occurring in 20\%-30\% of the patients [1]. Uveitis is inflammation in the middle layer of the eye, and it can affect the front (anterior) or the back (posterior) part of the eye. Uveitis can cause eye pain, sensitivity to light, and blurred vision, and it requires immediate medical attention [1]. 
\end{itemize}

\subsection{Is fever a common symptom of Jock Itch? (Ground Truth: No, AI: No)}

\medskip
\textbf{\textsc{Original response from Copilot in Bing (July 20, 2023)}}
\begin{quote}
No, fever is not a common symptom of Jock Itch. However, if there are signs of infection such as fever, chills, a pus-like discharge, or worsening pain, swelling, warmth, or redness, it is important to see a healthcare provider [1].

\medskip
[1] Jock Itch Blisters: What You Need to Know - Verywell Health. \url{https://www.verywellhealth.com/what-to-know-about-jock-itch-blisters-5199753}
\end{quote}

\medskip
\noindent \textbf{\textsc{AI responses used}}
\begin{itemize}[leftmargin=*]
    \item \textbf{Not uncertain}: No, fever is not a common symptom of Jock Itch. However, if there are signs of infection such as fever, chills, a pus-like discharge, or worsening pain, swelling, warmth, or redness, it is important to see a healthcare provider [1].
    \item \textbf{Uncertain in the first-person perspective}: \diff{I'm not certain, but I would guess} fever is not a common symptom of Jock Itch. However, if there are signs of infection such as fever, chills, a pus-like discharge, or worsening pain, swelling, warmth, or redness, it is important to see a healthcare provider [1].
    \item \textbf{Uncertain in the general perspective}: \diff{There is uncertainty, but it seems like} fever is not a common symptom of Jock Itch. However, if there are signs of infection such as fever, chills, a pus-like discharge, or worsening pain, swelling, warmth, or redness, it is important to see a healthcare provider [1].
\end{itemize}

\subsection{Can an adult who has not had chickenpox get shingles? (Ground Truth: No, AI: No)}

\medskip
\textbf{\textsc{Original response from Copilot in Bing (July 10, 2023)}}
\begin{quote}
No, an adult who has not had chickenpox cannot get shingles. Shingles is caused by the varicella-zoster virus which is the same virus that causes chickenpox. If you have never had chickenpox, you cannot get shingles. However, if you have not had chickenpox and you come into contact with someone who has shingles, you can get chickenpox [1, 2].

\medskip
[1] How Shingles Can Cause Chickenpox - Verywell Family. \url{https://www.verywellfamily.com/is-shingles-contagious-2634655}

[2] Can You Get Shingles If You Never Had Chickenpox? - Healthline. \url{https://www.healthline.com/health/can-you-get-shingles-if-you-never-had-chickenpox}
\end{quote}

\bigskip
\noindent \textbf{\textsc{AI responses used}}
\begin{itemize}[leftmargin=*]
    \item \textbf{Not uncertain}: No, an adult who has not had chickenpox cannot get shingles. Shingles is caused by the varicella-zoster virus which is the same virus that causes chickenpox. If you have never had chickenpox, you cannot get shingles. However, if you have not had chickenpox and you come into contact with someone who has shingles, you can get chickenpox [1] [2].
    \item \textbf{Uncertain in the first-person perspective}: \diff{I'm not sure, but it seems to me that} an adult who has not had chickenpox cannot get shingles. Shingles is caused by the varicella-zoster virus which is the same virus that causes chickenpox. If you have never had chickenpox, you cannot get shingles. However, if you have not had chickenpox and you come into contact with someone who has shingles, you can get chickenpox [1] [2].
    \item \textbf{Uncertain in the general perspective}: \diff{There is uncertainty, but it seems like} an adult who has not had chickenpox cannot get shingles. Shingles is caused by the varicella-zoster virus which is the same virus that causes chickenpox. If you have never had chickenpox, you cannot get shingles. However, if you have not had chickenpox and you come into contact with someone who has shingles, you can get chickenpox [1] [2].
\end{itemize}

\end{document}